\documentclass[journal]{IEEEtranTIE}
\usepackage{graphicx}
\usepackage{cite}
\usepackage{picinpar}
\usepackage{amsmath}
\usepackage{amssymb} 
\usepackage{flushend}
\usepackage[utf8]{inputenc}  
\usepackage{colortbl}
\usepackage{xcolor}
\usepackage{cuted}
\usepackage{soul}
\usepackage{multirow}
\usepackage{pifont}
\usepackage{color}
\usepackage{alltt}
\usepackage[hyphens]{xurl}
\usepackage[hidelinks]{hyperref}
\usepackage{enumerate}
\usepackage{siunitx}
\usepackage{epstopdf}
\usepackage{pbox}
\usepackage{etoolbox} 
\usepackage{balance}

\begin{document}
\title{	Impedance Modeling and Stability Analysis of Droop-Controlled Inverter Under Unbalanced \\Power Grid Operating Conditions}

\author{
	\vskip 1em
	
	Qiang~Zeng, 
	Lipeng~Zhu, \emph{Senior Member,~IEEE},
	Yang~Li,
    Yi Lei,
    Quan~Zhou, \emph{Senior Member,~IEEE},
    Jiayong~Li, \emph{Senior Member,~IEEE},
    Cong~Zhang, \emph{Senior Member,~IEEE},
    Bingxu~Li,
    Zhikang~Shuai, \emph{Senior Member,~IEEE}

	\thanks{This work was supported in part by the National Natural Science Foundation of China under Grants 52207094, 52125705, 52207202 and in part by the Science and Technology Innovation Program of Hunan Province under Grants 2025RC1023, 2024RC3110, and 2023RC3114. (\textit{Corresponding author: Lipeng Zhu})
	
	Qiang Zeng, Lipeng Zhu, Yang Li, Yi Lei, Quan Zhou, Jiayong Li, Cong Zhang, Bingxu Li and Zhikang Shuai are with the College of Electrical and Information Engineering, Hunan University, Changsha 410082, China (e-mail: zq0113@hnu.edu.cn; lpzhu@hnu.edu.cn; yangli2015@hnu.edu.cn; B2309S0389@hnu.edu.cn; zquan@hnu.edu.cn; jyli@hnu.edu.cn; zcong@hnu.edu.cn; bingxuli@hnu.edu.cn; szk@hnu.edu.cn).
}
}

\maketitle

\begin{abstract}
With the growing integration of renewable energy sources into power grids, the risks of oscillation caused by interactions between grid-tied inverters and the grids are becoming increasingly prominent. Although existing studies have made significant progress in inverter modeling and oscillatory stability analysis, most of them do not sufficiently consider complex mirror frequency coupling effects (MFCE) under unbalanced operating conditions, leading to unreliable models and erroneous stability analysis results. To address this inadequacy, this work develops a novel sequence impedance modeling scheme that can be widely applied to unbalanced operating conditions. In particular, taking a representative type of grid-forming inverter for instance, i.e., droop-controlled inverter (DCI), a single-input single-output sequence impedance modeling method based on harmonic linearization (HL) is proposed to comprehensively model both a given DCI and the connected grid. By accounting for multi-frequency interactions within the DCI, this method captures MFCE and unbalanced factors, leading to a more accurate impedance model. Further, the dominant factors influencing system stability are identified with a combination of normalized sensitivity analysis and proportional weighting. Finally, the detailed impacts of these dominant factors on system stability margin under three typical unbalanced operating conditions are analyzed through the Bode criterion. The effectiveness and reliability of the whole scheme proposed in this work are validated on the constructed grid-connected droop-controlled experimental platform.
\end{abstract}


\markboth{IEEE TRANSACTIONS ON INDUSTRIAL ELECTRONICS}%
{}

\definecolor{limegreen}{rgb}{0.2, 0.8, 0.2}
\definecolor{forestgreen}{rgb}{0.13, 0.55, 0.13}
\definecolor{greenhtml}{rgb}{0.0, 0.5, 0.0}

\section{Introduction}
\label{sec:introduction}
\IEEEPARstart{W}{ith} the rapid development of renewable energy resources, the penetration of power electronics-enabled inverters into modern power grids has been increasing significantly. However, interactions between the inverters and the power grids pose a high risk of inducing sub/super-synchronous oscillations \cite{lin2021improving}, which have become more and more pronounced in various regions worldwide, such as Ajo \cite{adams2012ercot} and Minnesota \cite{narendra2011new} in the USA, as well as Guyuan \cite{liu2017stability} and Hami \cite{liu2018oscillatory} in China. Such interactions may cause large-scale tripping of renewable energy units from power grids \cite{wang2023partition}, posing significant challenges to the stability of the grids \cite{11488592,hu2022impedance}. In microgrids, three-phase droop-controlled inverters (DCIs) are one of the commonly adopted grid-forming power interfaces to accommodate distributed generation. If not properly controlled, DCIs could also introduce oscillatory stability issues. To mitigate potential instability risks and ensure the stable operation of distributed renewable energy systems, it is imperative to reliably perform wideband frequency dynamics modeling and stability analysis for DCIs and the grids.

Regarding the modeling of wideband frequency dynamics for DCIs, there are two primary approaches: state-space methods and impedance-based methods \cite{li2022systematic}. Existing studies have widely employed the state-space methods for modeling and stability analysis of DCIs \cite{deng2021analysis,liu2015comparison,li2017small}. However, the state-space model must be re-established when system operating conditions or parameters change \cite{li2025sequence}, and its complexity increases rapidly for multi-converter parallel systems. In contrast, the impedance-based method divides the grid-connected renewable system into two independent subsystems, only the changed subsystem needs to be re-modeled. Additionally, in multi-converter parallel systems, the impedance-based method enables independent modeling of each converter and their systematic combination, thereby reducing modeling complexity. Therefore, the impedance-based method is more practical for real-world applications. Impedance-based methods can be generally classified into two major categories: dq impedance modeling method and sequence impedance modeling method \cite{liu2018oscillatory}. The dq and sequence impedance models are both built upon linearization techniques, but have different scopes of application. Specifically, the conventional dq impedance modeling approach is not suitable for a three-phase unbalanced system, since the system no longer exhibits linear time-invariant behavior in the dq reference frame under unbalanced conditions, thereby violating the modeling assumption of the dq impedance modeling approach \cite{wang2019small,wang2020stability,guo2024virtual}. In contrast, the sequence impedance modeling method has better applicability, as it is effective for inverter modeling under both balanced and unbalanced operating conditions. Given the broad applicability, this paper primarily focuses on sequence impedance-based oscillatory behavior modeling of DCIs.

In terms of sequence impedance-based DCI modeling, considerable research efforts have been made in the research community. For example, positive- and negative-sequence impedance models of a given DCI are derived in \cite{zhao2024impedance} based on harmonic linearization (HL) theory, and system stability is analyzed using the Nyquist criterion. Similarly, a sequence impedance model for inverters with droop-based multi-loop control is developed in \cite{yan2021sequence}. To achieve a unified sequence impedance formulation, Elshenawy \textit{et al}. \cite{elshenawy2023unified} develop an HL-based sequence impedance model for synchronous generators and DCIs. The sequence impedance models in \cite{zhao2024impedance,yan2021sequence,elshenawy2023unified} are derived under the assumption of three-phase balanced operating conditions. However, DCIs frequently operate under unbalanced conditions in practice due to unequal load variations among phases. In such cases, both voltage and current contain fundamental negative-sequence components, which are neglected in \cite{zhao2024impedance,yan2021sequence,elshenawy2023unified}. Besides, the mirror frequency coupling effect (MFCE) within DCIs tends to be more pronounced under unbalanced conditions, and it significantly affects impedance characteristics. However, it is not considered in \cite{zhao2024impedance,yan2021sequence,elshenawy2023unified}. As a result, the impedance models built in \cite{zhao2024impedance,yan2021sequence,elshenawy2023unified} may not accurately capture the actual oscillatory behaviors of DCIs under unbalanced operating conditions, which may lead to erroneous stability analysis results.

In recent years, some researchers have attempted to develop two-dimensional sequence impedance matrix models for DCIs by accounting for MFCE \cite{rong2023modeling,elshenawy2024analysis,hu2019sequence}. Compared with the aforementioned studies \cite{zhao2024impedance,yan2021sequence,elshenawy2023unified}, these works consider the coupling between positive- and negative-sequence impedances, resulting in more accurate modeling of DCIs' impedance characteristics. Nonetheless, the impedance matrix models in \cite{rong2023modeling,elshenawy2024analysis,hu2019sequence} are still built upon the assumption of three-phase balanced conditions. Considering the MFCE and unbalanced conditions, Guo \textit{et al}. \cite{guo2022harmonic} derive a harmonic transfer function model in the \(\alpha\beta\) frame for DCIs in islanded microgrids with unbalanced loads. However, compared with the sequence impedance model, the order of the harmonic transfer function model is relatively high, resulting in a heavy computational burden. Moreover, since the dynamic characteristics of the grid side are not considered in the modeling process, the developed model may not be suitable for analysis under grid-connected operating conditions. Overall, considering the ubiquitous unbalanced conditions in practice, how to comprehensively model grid-connected DCIs'  impedance characteristics towards oscillatory stability analysis with sufficient consideration of the MFCE remains an unsolved challenge.

Given the aforementioned research gap, this paper presents a comprehensive impedance modeling scheme for DCIs under unbalanced conditions and conducts stability analysis by using the Bode criterion. Unlike most existing studies that neglect the MFCE or unbalanced operating conditions, the scheme proposed in this paper aims to accurately capture the internal dynamic behavior of the DCI and quantitatively assess the impact of variations in both key parameters and the degree of unbalance on the stability of the grid-connected system under three typical unbalanced scenarios. First, a single-input single-output (SISO) sequence impedance modeling approach based on the HL method is introduced to precisely capture the internal dynamic behavior of the DCI under unbalanced conditions. Next, the accuracy of the developed DCI model is compared with that of existing models. Furthermore, the influence of grid voltage unbalance level on MFCE and the impedance characteristics of the DCI are investigated. Finally, the key parameters that significantly affect the stability of the grid-connected system are identified via a sensitivity-based quantitative analysis method, and their specific effects on the system's stability margin under unbalanced scenarios are analyzed, providing practical guidance for parameter optimization and stability enhancement. The proposed modeling framework, including the modeling steps and derivation logic, can adapt well to other grid-forming inverters to establish their impedance models according to their specific topologies. Also, the stability analysis technique introduced in this paper can be used to evaluate their stability status. The major contributions and merits of this work are as follows.
\begin{enumerate}
\item{This work develops a high-precision SISO impedance modeling scheme for DCIs to enhance the reliability of stability analysis in grid-connected renewable generation systems. Compared with existing DCI models,} the proposed scheme achieves better applicability in practical unbalanced operating conditions by fully accounting for the MFCE and fundamental negative-sequence components.
\item{By introducing a comprehensive frequency coupling coefficient with a concise computational formulation and an intuitive representation of MFCE strength, this paper facilitates clear identification of the dominant frequency ranges in which variations in grid voltage unbalance influence the MFCE.}
\item{By integrating normalized sensitivity analysis with proportional weighting, a systematic stability analysis approach is adopted to identify the dominant parameters that contribute to instability in grid-connected systems. The influence of these parameters on the system stability margin under three typical unbalanced operating conditions is analyzed using the Bode criterion. This provides practical guidance for optimizing control parameters to enhance system stability.}
\end{enumerate}

\section{Proposed Impedance Modeling Scheme  and Impedance Characteristic Analysis}
This section utilizes the HL method to develop an impedance modeling scheme for the main and control circuits of DCI. The proposed scheme rigorously incorporates the effects of MFCE and unbalanced operating conditions to ensure accurate impedance modeling. In addition, the influence of the grid voltage unbalanced degree on MFCE and the impedance characteristics is investigated.

\subsection{Overview of the Harmonic Linearization Method}
The HL method constructs a small-signal linear model along a periodic operating trajectory. Mathematically, this is achieved by injecting a harmonic perturbation at the point of common coupling (PCC) and analyzing the resulting small-signal responses of the system variables to extract the components at the perturbation and coupling frequencies.

To derive the impedance of a grid-connected converter using the HL method, the time-domain variables are transformed into frequency-domain. This methodology is firmly rooted in the principles of Fourier series expansion and the convolution theorem \cite{sun2009input}. The HL modeling procedure can be summarized in the following steps:

Step 1: Formulate the main circuit equation of the grid-connected converter in the frequency domain, which characterizes the relationship between the bridge-arm midpoint voltage and the PCC voltage and current.

Step 2: Assume small-signal perturbation injection and transform the PCC voltage and current into the frequency domain via Fourier series expansion.

Step 3: Drive the frequency-domain expressions of the electrical quantities associated with the converter control system using the convolution theorem. Then, derive the frequency-domain relationship between the bridge-arm midpoint voltage and the perturbation voltage and current.

Step 4: Substitute the expression of the bridge-arm midpoint voltage into the main circuit equation in Step 1 to establish the frequency-domain voltage-current relationship and finally derive the converter impedance under unbalanced conditions.

\subsection{Harmonic Linearization Method-Based Impedance Modeling of the Droop Inverter}

1) Modeling of the main circuit

According to Fig. 1, the frequency-domain expressions of mid-point voltages, PCC voltages, and PCC currents are derived as follows:

\begin{equation}
\label{eq1}
\begin{bmatrix}
    e_\mathrm{a} \\ e_\mathrm{b} \\ e_\mathrm{c}
\end{bmatrix}
= s L_\mathrm{f}
\begin{bmatrix}
    i_\mathrm{a} \\ i_\mathrm{b} \\ i_\mathrm{c}
\end{bmatrix}
+ \left( 1 + s^2 L_\mathrm{f} C_\mathrm{f} \right)
\begin{bmatrix}
    u_\mathrm{a} \\ u_\mathrm{b} \\ u_\mathrm{c}
\end{bmatrix}
\end{equation}%
where $e_\mathrm{a}$, $e_\mathrm{b}$, $e_\mathrm{c}$ denote the mid-point voltages of the DCI; $L_\mathrm{f}$, $C_\mathrm{f}$ represent the inverter-side inductance and capacitor, respectively; $u_\mathrm{a}$, $u_\mathrm{b}$, $u_\mathrm{c}$ are the three-phase PCC voltages; $i_\mathrm{a}$, $i_\mathrm{b}$, $i_\mathrm{c}$ are the three-phase PCC currents.

\begin{figure}
    \centering
    \includegraphics[width=1\linewidth]{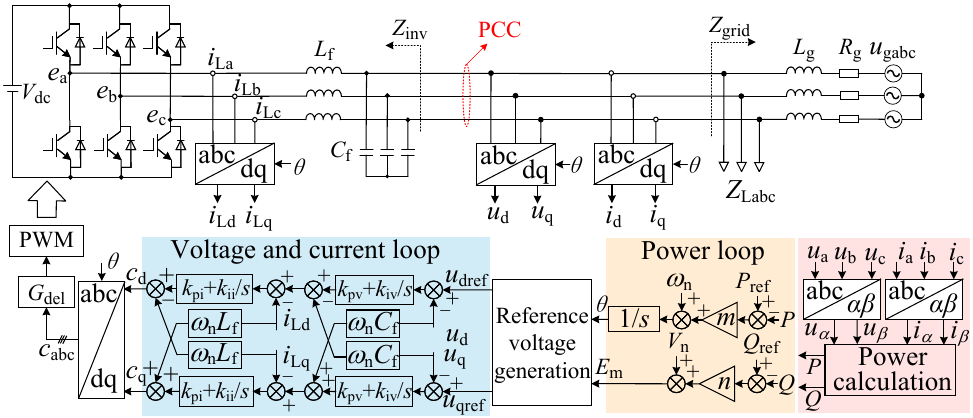}
    \caption{Structural diagram of the droop-controlled inverter and the power grid.}
    \label{fig.1}
\end{figure}
When a small amplitude positive-sequence voltage perturbation at a frequency $f_\mathrm{p}$ is imposed at the PCC, \(f_\mathrm{p}\) is the disturbance frequency. According to the basic principle of frequency coupling, it will induce a positive-sequence response current at frequency $f_\mathrm{p}$ and a negative-sequence coupling current response at frequency \(f_{\mathrm{p1}}\) (\( f_{\mathrm{p1}} = f_{\mathrm{p}} - 2f_{\mathrm{1}} \)), where \(f_\mathrm{p1}\) is the coupling frequency, \(f_1\) is the fundamental frequency. Due to the grid impedance, the negative-sequence coupling current generates a negative-sequence voltage response at frequency \(f_\mathrm{p1}\) \cite{wang2022admittance}. Since the amplitude of the injected perturbation is small, the resulting disturbance and coupling frequency components are not regarded as significant contributors to system imbalance. In most practical scenarios, system imbalance is primarily caused by asymmetric loads, grid impedance, and grid voltage. Overall, when a grid-connected system operates under unbalanced conditions, the PCC voltage and current contain a fundamental negative-sequence component at the frequency $f_{\mathrm{1}}$. Consequently, in the time domain, the PCC voltage and current of phase A are expressed as follows:
\begin{align}
\label{eq2}
u_{\mathrm{a}}(t) &= V_1 \cos(2\pi f_1 t) + V_2 \cos(2\pi f_1 t + \varphi_{\mathrm{v2}}) +  \notag \\
&\quad V_{\mathrm{p}} \cos(2\pi f_{\mathrm{p}} t + \varphi_{\mathrm{vp}}) + V_{\mathrm{p1}} \cos(2\pi f_{\mathrm{p1}} t + \varphi_{\mathrm{vp1}})
\end{align}
\begin{align}
\label{eq3}
i_{\mathrm{a}}(t) &= I_1 \cos(2\pi f_1 t + \varphi_{\mathrm{i1}}) + I_2 \cos(2\pi f_1 t + \varphi_{\mathrm{i2}}) + \notag \\
&\quad I_{\mathrm{p}} \cos(2\pi f_{\mathrm{p}} t + \varphi_{\mathrm{ip}}) + I_{\mathrm{p1}} \cos(2\pi f_{\mathrm{p1}} t + \varphi_{\mathrm{ip1}})
\end{align}
where $V_\mathrm{1}$, $V_\mathrm{2}$, $V_\mathrm{p}$ and $V_\mathrm{p1}$ denote the amplitudes of the fundamental positive-sequence voltage, fundamental negative-sequence voltage, disturbance voltage at $f_\mathrm{p}$, and coupling voltage at $f_\mathrm{p1}$, respectively; $I_\mathrm{1}$, $I_\mathrm{2}$, $I_\mathrm{p}$ and $I_\mathrm{p1}$ denote the corresponding current amplitudes; $\varphi_\mathrm{v2}$, $\varphi_\mathrm{vp}$, $\varphi_\mathrm{vp1}$, $\varphi_\mathrm{i1}$, $\varphi_\mathrm{i2}$, $\varphi_\mathrm{ip}$, $\varphi_\mathrm{ip1}$ are the initial phase angles of the corresponding voltage and current components. 

In the modeling process, coupling terms such as \(f_\text{p}-f_1\), \(f_\text{p}+f_1\), and \(f_\text{p}+2f_1\) are omitted due to their negligible amplitudes \cite{xu2022modeling}. According to the principle of Fourier series expansion \cite{website}, Eqs. (\ref{eq2}) and (\ref{eq3}) can be expressed in the frequency-domain as
\begin{equation}
\label{eq4}
u_{\mathrm{a}}[f] =
\begin{cases}
    \mathbf{V}_1, & f = \pm f_1 \\
    \mathbf{V}_2, & f = \pm f_1 \\
    \mathbf{V}_{\mathrm{p}}, & f = \pm f_{\mathrm{p}} \\
    \mathbf{V}_{\mathrm{p1}}, & f = \pm f_{\mathrm{p1}}
\end{cases}
i_{\mathrm{a}}[f] =
\begin{cases}
    \mathbf{I}_1, & f = \pm f_1 \\
    \mathbf{I}_2, & f = \pm f_1 \\
    \mathbf{I}_{\mathrm{p}}, & f = \pm f_{\mathrm{p}} \\
    \mathbf{I}_{\mathrm{p1}}, & f = \pm f_{\mathrm{p1}}
\end{cases}
\end{equation}
where \( \mathbf{V}_1\)=\(V_1/2\), \( \mathbf{V}_2\)=\((V_2 / 2) e^{\pm j \varphi_{\mathrm{v2}}}\), \( \mathbf{V}_{\text{p}}\)=\((V_{\text{p}} / 2) e^{\pm j \varphi_{\mathrm{vp}}}\), \( \mathbf{V}_{\text{p1}}\)=\((V_{\text{p1}} / 2) e^{\pm j \varphi_{\mathrm{vp1}}}\), \( \mathbf{I}_1\)=\((I_1 / 2) e^{\pm j \varphi_{\mathrm{i1}}}\),
\( \mathbf{I}_2\)=\((I_2 / 2) e^{\pm j \varphi_{\mathrm{i2}}}\), \( \mathbf{I}_{\text{p}}\)=\((I_{\text{p}} / 2) e^{\pm j \varphi_{\mathrm{ip}}}\), \( \mathbf{I}_{\text{p1}}\)=\((I_{\text{p1}} / 2) e^{\pm j \varphi_{\mathrm{ip1}}}\).

2) Modeling of the control circuit

According to Eq. (\ref{eq4}), the frequency-domain expressions of the DCI's PCC voltage and current in the $\alpha$$\beta$ frame are given as follows:
\begin{equation}
\label{eq444}
u_{\mathrm{\alpha}}[f] =
\begin{cases}
    \mathbf{V}_1, & f = \pm f_1 \\
    \mathbf{V}_2, & f = \pm f_1 \\
    \mathbf{V}_{\mathrm{p}}, & f = \pm f_{\mathrm{p}} \\
    \mathbf{V}_{\mathrm{p1}}, & f = \pm f_{\mathrm{p1}}
\end{cases}
i_{\mathrm{\alpha}}[f] =
\begin{cases}
    \mathbf{I}_1, & f = \pm f_1 \\
    \mathbf{I}_2, & f = \pm f_1 \\
    \mathbf{I}_{\mathrm{p}}, & f = \pm f_{\mathrm{p}} \\
    \mathbf{I}_{\mathrm{p1}}, & f = \pm f_{\mathrm{p1}}
\end{cases}
\end{equation}

\begin{equation}
\label{eq454}
u_{\mathrm{\beta}}[f] =
\begin{cases}
    \mp j\mathbf{V}_1, & f = \pm f_1 \\
    \pm j\mathbf{V}_2, & f = \pm f_1 \\
    \mp j\mathbf{V}_{\mathrm{p}}, & f = \pm f_{\mathrm{p}} \\
    \pm j\mathbf{V}_{\mathrm{p1}}, & f = \pm f_{\mathrm{p1}}
\end{cases}
i_{\mathrm{\beta}}[f] =
\begin{cases}
    \mp j\mathbf{I}_1, & f = \pm f_1 \\
    \pm j\mathbf{I}_2, & f = \pm f_1 \\
    \mp j\mathbf{I}_{\mathrm{p}}, & f = \pm f_{\mathrm{p}} \\
    \pm j\mathbf{I}_{\mathrm{p1}}, & f = \pm f_{\mathrm{p1}}
\end{cases}
\end{equation}

In the amplitude invariant form, the time-domain expressions of the active power \(P\) and reactive power \(Q\) of the DCI are related as follows \cite{elshenawy2024analysis}:
\begin{equation}
\label{eq7}
\begin{cases}
    P = 1.5 G_f(s)\left( u_{\alpha} i_{\alpha} + u_{\beta} i_{\beta} \right) \\
    Q = 1.5 G_f(s)\left( u_{\beta} i_{\alpha} - u_{\alpha} i_{\beta} \right)
\end{cases}
\end{equation}
where \(G_f(s)\) is the transfer function of the low-pass filter, \(G_f(s)=\omega_c/(s+\omega_c)\), \(\omega_c\) is the cutoff angular frequency.

According to Eq. (\ref{eq7}) and the frequency-domain convolution theorem \cite{website}, the frequency-domain expressions of \(P\) and \(Q\) can be obtained, as shown in Eqs. (\ref{eq8}) and (\ref{eq9}).
\begin{equation}
\label{eq8}
P[f] =
\begin{cases}
    3 G_f(s)\big( \mathbf{V}_1 \mathbf{I}_1^* + \mathbf{I}_1 \mathbf{V}_1^* 
    + \\ \quad \mathbf{V}_2 \mathbf{I}_2^* + \mathbf{V}_2^* \mathbf{I}_2 \big), 
    & f = \text{dc} \\[4pt]
    
    3 G_f(s) \big( \mathbf{V}_{\mathrm{p}} \mathbf{I}_1^* + \mathbf{I}_{\mathrm{p}} \mathbf{V}_1^* 
    + \\ \quad \mathbf{V}_{\mathrm{p1}} \mathbf{I}_1 + \mathbf{V}_1 \mathbf{I}_{\mathrm{p1}} \big), 
    & f = \pm (f_{\text{p}} - f_1) \\[4pt]
    
    3 G_f(s) \big( \mathbf{V}_1 \mathbf{I}_2 + \mathbf{V}_2 \mathbf{I}_1 \big), 
    & f = \pm 2 f_1
\end{cases}
\end{equation}

\begin{equation}
\label{eq9}
Q[f] =
\begin{cases}
    \pm 3j G_f(s)\big( -\mathbf{V}_1 \mathbf{I}_1^* + \mathbf{V}_1^* \mathbf{I}_1 
    + \\ \quad \mathbf{V}_2 \mathbf{I}_2^* - \mathbf{V}_2^* \mathbf{I}_2 \big), 
    & f = \text{dc} \\[4pt]
    \pm 3j G_f(s)\big( -\mathbf{V}_{\mathrm{p}} \mathbf{I}_1^* + \mathbf{V}_1^* \mathbf{I}_{\mathrm{p}} 
    + \\ \quad \mathbf{V}_{\mathrm{p1}} \mathbf{I}_1 - \mathbf{V}_1 \mathbf{I}_{\mathrm{p1}} \big), 
    & f = \pm (f_{\text{p}} - f_1) \\[4pt]
    \pm 3j G_f(s) \big( -\mathbf{V}_1 \mathbf{I}_2 + \mathbf{V}_2 \mathbf{I}_1 \big), 
    & f = \pm 2 f_1
\end{cases} \
\end{equation}
where ``*" represents the complex conjugate operator.

According to Fig. 1, the mathematical equation of the active power control loop is as follows:

\begin{equation}
\theta = \left( m \left( P_{\mathrm{ref}} - P \right) + \omega_n \right) / s
\label{eq:theta}
\end{equation}
where  $\theta$ is the modulation wave phase angle; $m$ is the active power-frequency droop coefficient; $P_\mathrm{ref}$ is the active power reference; $\omega_\mathrm{n}$ is the rated angular frequency of the grid. 

By substituting Eq. (\ref{eq8}) into Eq. (\ref{eq:theta}), the frequency-domain expression of \(\theta\) is obtained as:
\begin{equation}
\label{eq10}
\theta[f] =
\begin{cases}
    (m P_{\text{ref}} + \omega_{\text{n}})/s + K(s) P_0(s), & f = \text{dc} \\[4pt]
    K(s) P_1(s), & f = \pm (f_{\text{p}} - f_1) \\[4pt]
    K(s) P_2(s), & f = \pm 2f_1
\end{cases}
\end{equation}
where \(K(s)=-m/s\); \(P_0(s)=3G_f(s)(\mathbf{V}_1 \mathbf{I}_1^*+\mathbf{I}_1 \mathbf{V}_1^*+\mathbf{V}_2 \mathbf{I}_2^*+\mathbf{V}_2^*\mathbf{I}_2\); \(P_1(s)=3G_f(s)(\mathbf{V}_{\text{p}} \mathbf{I}_1^*+\mathbf{I}_{\text{p}} \mathbf{V}_1^*+\mathbf{V}_{\text{p1}} \mathbf{I}_1+\mathbf{V}_1 \mathbf{I}_{\text{p1}})\); \(P_2(s)=3G_f(s)(\mathbf{V}_1 \mathbf{I}_2+\mathbf{V}_2 \mathbf{I}_1).\)
    
After applying a positive-sequence perturbation voltage at the PCC, the modulation wave phase angle exhibits a phase perturbation \( \Delta\theta \) \cite{elshenawy2023unified}, i.e., \(\theta=\theta_1+\Delta\theta\), \(\theta_1\) is the fundamental component of the \(\theta\). Based on Eq. (\ref{eq10}), the expression of \( \Delta\theta \) can be obtained as follows:
\begin{equation}
\label{eq11}
\Delta \theta [f] =
\begin{cases}
    K(s) P_1(s), & f = \pm (f_{\text{p}} - f_1) \\
    K(s) P_2(s), & f = \pm 2f_1
\end{cases}
\end{equation}

According to Fig. 1, the mathematical equation of the reactive power control loop is derived as follows:

\begin{equation}
\label{eq1124}
E_{\text{m}} =n(Q_{\text{ref}}-Q)+V_{\text{n}}
\end{equation}where \(E_{\text{m}}\) is the output modulation wave voltage amplitude of the DCI; $n$ represents the reactive power-voltage droop coefficient; $Q_\mathrm{ref}$ represents the reactive power reference; $V_\mathrm{n}$ represents the grid voltage reference.

By substituting Eq. (\ref{eq9}) into Eq. (\ref{eq1124}), the frequency-domain expression of \(E_{\text{m}}\) is as follows:

\begin{equation}
\label{eq12}
E_{\text{m}}[f] =
\begin{cases}
    n (Q_{\text{ref}} - Q_0(s)) + V_{\text{n}}, & f = \text{dc} \\
    -n Q_1(s), & f = \pm (f_{\text{p}} - f_1) \\
    -n Q_2(s), & f = \pm 2f_1
\end{cases}
\end{equation}
where \(Q_0(s)=\pm 3j G_f(s)\big( -\mathbf{V}_1 \mathbf{I}_1^* + \mathbf{V}_1^* \mathbf{I}_1 
    +\mathbf{V}_2 \mathbf{I}_2^* - \mathbf{V}_2^* \mathbf{I}_2 \big)\); \(Q_1(s)= \pm 3j G_f(s)\big( -\mathbf{V}_{\mathrm{p}} \mathbf{I}_1^* + \mathbf{V}_1^* \mathbf{I}_{\mathrm{p}} 
    + \mathbf{V}_{\mathrm{p1}} \mathbf{I}_1 - \mathbf{V}_1 \mathbf{I}_{\mathrm{p1}} \big)\); \(Q_2(s)= \pm 3j G_f(s) \big( -\mathbf{V}_1 \mathbf{I}_2 + \mathbf{V}_2 \mathbf{I}_1 \big)\).

Based on Eq. (\ref{eq12}), the frequency-domain expressions of the d-axis and q-axis reference voltages in the voltage loop can be formulated as follows:
\begin{equation}
\label{eq13}
u_{\text{dref}}[f] =
\begin{cases}
    n (Q_{\text{ref}} - Q_0(s)) + V_{\text{n}}, & f = \text{dc} \\
    -n Q_1(s), & f = \pm (f_{\text{p}} - f_1) \\
    -n Q_2(s), & f = \pm 2f_1
\end{cases}
\end{equation}

\begin{equation}
\label{eq14}
u_{\text{qref}}[f] = 0
\end{equation} 
where $u_\mathrm{dref}$ and $u_\mathrm{qref}$ denote the d-axis and q-axis voltage reference values in the voltage loop, respectively.

As illustrated in Fig. 1, the Park transformation is utilized to convert the three-phase voltages and currents into the dq-frame. The Park transformation matrix \(T_{abc/dq}(\theta)\), accounting for small perturbations in the phase angles, is given by:

\begin{flalign}
& T_{abc/dq}(\theta) =
\frac{2}{3}
\begin{bmatrix}
\cos(\Delta\theta) & \sin(\Delta\theta) \\
-\sin(\Delta\theta) & \cos(\Delta\theta)
\end{bmatrix}
\notag \\   
& \cdot
\begin{bmatrix}
\cos\theta_1 & \cos\left(\theta_1 - 2\pi/3\right) & \cos\left(\theta_1 + 2\pi/3\right) \\
-\sin\theta_1 & -\sin\left(\theta_1 - 2\pi/3\right) & -\sin\left(\theta_1 + 2\pi/3\right)
\end{bmatrix}
\label{eq:Tdq}   
\end{flalign}

The three-phase voltages \(u_{\text{abc}}\) at PCC and filter inductor currents \(i_{\text{Labc}}\) are transformed by Park transformation to obtain the dq axis voltages and filter inductor currents when phase angle perturbation is considered: 

\begin{equation}
\label{eq:ud}
u_{\mathrm{d}}[f] =
\begin{cases}
V_1, & f = \mathrm{dc} \\[6pt]
\mathbf{V}_{\mathrm{p}} + \mathbf{V}_{\mathrm{p1}}, & f = \pm \left( f_{\mathrm{p}} - f_1 \right) \\[6pt]
\mathbf{V}_2, & f = \pm 2f_1
\end{cases}
\end{equation}
\begin{equation}
\label{eq:uq}
u_{\mathrm{q}}[f] =
\begin{cases}
0, & f = \mathrm{dc} \\[8pt]
- K(s) P_{1}(s) V_{1} \mp j\mathbf{V}_{\mathrm{p}} \pm j\mathbf{V}_{\mathrm{p1}}, 
& f = \pm \left( f_{\mathrm{p}} - f_{1} \right) \\[8pt]
- K(s) P_{2}(s) V_{1} \pm j\mathbf{V}_{2}, 
& f = \pm 2f_{1}
\end{cases}
\end{equation}
\begin{equation}
\label{eq:iLd}
i_{\mathrm{L_d}}[f] =
\begin{cases}
A_0(s),
& f = \mathrm{dc} \\
\begin{aligned}
A_1(s),
\end{aligned}
& f = \pm\left(f_{\mathrm{p}} - f_{1}\right) \\
\begin{aligned}
A_2(s),
\end{aligned}
& f = \pm 2f_{2}
\end{cases}
\end{equation}
\begin{equation}
\label{eq:iLq}
i_{\text{Lq}}[f] =
\begin{cases}
B_0, 
& f=\mathrm{dc} \\
\begin{aligned}
B_1(s),
\end{aligned}
& f=\pm\!\left(f_p-f_1\right) \\
\begin{aligned}
B_2(s),
\end{aligned}
& f=\pm 2f_1
\end{cases}
\end{equation}
where \(u_{\text{d}}\) and \(u_{\text{q}}\) are the d-axis and q-axis voltage at PCC, respectively; \(i_{\text{Ld}}\) and \(i_{\text{Lq}}\) are the d-axis and q-axis current of the filter inductor, respectively; \(Y_{\mathrm{c}}(s)=sC_{\mathrm{f}}\); \(A_0(s)=Y_{\mathrm{c}}(s)V_{1} + I_{1}\cos\varphi_{\text{i1}}\); \(A_1(s)=Y_{\mathrm{c}}(s)\!\left(\mathbf{V}_{\mathrm{p}} + \mathbf{V}_{\mathrm{p1}}\right)+ \mathbf{I}_{\mathrm{p}} + \mathbf{I}_{\mathrm{p1}}+ K(s)P_{1}(s)I_{1}\sin\varphi_{\text{i1}}\); \(A_2(s)= Y_{\mathrm{c}}(s)\mathbf{V}_{2} + \mathbf{I}_{2}+K(s)P_{2}(s)I_{1}\sin\varphi_{\text{i1}}\); \(B_0= I_1\sin\varphi_{\text{i1}}\); \(B_1(s)=-K(s)P_{1}(s)(Y_{\mathrm{c}}(s)V_1+I_1\cos\varphi_{\text{i1}})+Y_{\mathrm{c}}(s)\bigl(\mp j \mathbf{V}_{\text{p}} \pm j \mathbf{V}_{\text{p1}}\bigr)\mp j \mathbf{I}_{\text{p}} \pm j \mathbf{I}_{\text{p1}}\); \(B_2(s)=-K(s)P_{2}(s)(Y_{\mathrm{c}}(s)V_1+I_1\cos\varphi_{\text{i1}})\pm j Y_{\mathrm{c}}(s)\mathbf{V}_2 \pm j \mathbf{I}_2\).

According to  the voltage loop configuration illustrated in Fig. 1 and Eqs. (\ref{eq13}), (\ref{eq14}), (\ref{eq:ud}), (\ref{eq:uq}), the frequency-domain expressions of \(I_{\text{dref}}\) and \(I_{\text{qref}}\) are obtained as follows.
\begin{equation}
\label{eq15}
I_{\text{dref}}[f] =
\begin{cases}
    C_0(s), & f = \text{dc} \\
    C_1(s), & f = \pm (f_{\text{p}} - f_1) \\
    C_2(s), & f = \pm 2f_1
\end{cases}
\end{equation}
\begin{equation}
\label{eq16}
I_{\text{qref}}[f] =
\begin{cases}
    D_0, & f = \text{dc} \\
    D_1(s), & f = \pm (f_{\text{p}} - f_1) \\
    D_2(s), & f = \pm 2f_1
\end{cases}
\end{equation}
where $I_\mathrm{dref}$ and $I_\mathrm{qref}$ denote the d-axis and q-axis current reference values of the current loop, respectively; \(C_0(s)=G_{\text{v}}(s)(n(Q_{\text{ref}}-Q_0(s))+V_{\text{n}}-V_1)\); \(C_1(s)=G_{\text{v}}(s)(-nQ_1(s)-\mathbf{V}_{\text{p}}-\mathbf{V}_{\text{p1}})-\omega_1 C_{\text{f}}(-K(s)P_1(s)V_1 \mp j \mathbf{V}_{\text{p}} \pm j \mathbf{V}_{\text{p1}}\); \(C_2(s)=G_{\text{v}}(s)(-nQ_2(s)-\mathbf{V}_2)-\omega_1 C_{\text{f}}(-K(s)P_2(s)V_1 \pm j \mathbf{V}_2)\); \(D_0=\omega_1 C_{\text{f}} V_1\); \(D_1(s)=G_{\text{v}}(s)(K(s)P_1(s)V_1 \pm j \mathbf{V}_{\text{p}} \mp j \mathbf{V}_{\text{p1}})+ \omega_1 C_{\text{f}}(\mathbf{V}_{\text{p}}+\mathbf{V}_{\text{p1}})\); \(D_2(s)=G_{\text{v}}(s)(K(s)P_2(s)V_1 \mp j \mathbf{V}_2)+\omega_1C_{\text{f}} \mathbf{V}_2\); \(G_{\text{v}}(s)=k_{\text{pv}}+k_{\text{iv}}/s\), \(G_{\text{v}}(s)\) is the transfer function of the voltage controllers, \(k_{\text{pv}}\) and \(k_{\text{iv}}\) are the proportional and integral gains of the voltage loop, respectively.

Based on the current loop configuration illustrated in Fig. 1 and Eqs. (\ref{eq:iLd}), (\ref{eq:iLq}), (\ref{eq15}), (\ref{eq16}), the frequency-domain expressions of \(c_{\text{dref}}\) and \(c_{\text{qref}}\) are derived as follows

\begin{equation}
\label{eq17}
c_{\text{d}}[f] =
\begin{cases}
    c_{\text{d0}}-\omega_1 L_{\text{f}} I_{\text{L}} \sin \varphi_{\text{iL}}, & f = \text{dc} \\
    G_{\text{i}}(s) \big(C_1(s) - A_1(s)\big)  \\
    - \omega_1 L_{\text{f}} B_1(s), & f = \pm (f_{\text{p}} - f_1) \\
    G_{\text{i}}(s) \big(C_2(s) - A_2(s)\big)  \\
    - \omega_{\text{1}} L_{\text{f}} B_2(s), & f = \pm 2f_1
\end{cases}
\end{equation}
\begin{equation}
\label{eq18}
c_{\text{q}}[f] =
\begin{cases}
    c_{\text{q0}} + \omega_1 L_{\text{f}} I_{\text{L}} \cos \varphi_{\text{iL}}, & f = \text{dc} \\
    G_{\text{i}}(s) \big(D_1(s) - B_1(s)\big)  \\
    +\omega_{\text{1}} L_{\text{f}} A_1(s), & f = \pm (f_{\text{p}} - f_1) \\
    G_{\text{i}}(s) \big(D_2(s) - B_2(s)\big) \\
    +\omega_{\text{1}} L_{\text{f}} A_2(s), & f = \pm 2f_1
\end{cases}
\end{equation}
where $c_\mathrm{d}$ and $c_\mathrm{q}$ are the modulation signals in the synchronous rotating reference frame, respectively; \(c_{\text{d0}}\) and \(c_{\text{q0}}\) represent DC components of $c_\mathrm{d}$ and $c_\mathrm{q}$, the detailed expressions are provided in \cite{zhao2024impedance}; \(G_{\text{i}}(s)=k_{\text{pi}} + {k_{\text{ii}}}/{s}\) is the transfer function of the current controller, $k_\mathrm{pi}$ and $k_\mathrm{ii}$ are the proportional and integral gains of the current loop, respectively; \(I_{\text{L}}\) represents the amplitude of the inductor current, while \(\varphi_{\text{iL}}\) is the initial phase angle of \(I_{\text{L}}\); \(\omega_{\text{1}}\) is defined as \(\omega_{\text{1}}=2 \pi f_{\text{1}}\).

According to Park's inverse transformation, the phase A modulation signals \( c_{\text{a}}[\pm f_{\text{p}}] \) and \( c_{\text{a}}[\pm f_{\text{p1}}] \) can be obtained, their detailed expressions are provided in a downloadable supplementary file \cite{website} [see Eqn. (A6)--(A7) in the file]. Thus, the frequency-domain expression of the mid-point voltage of phase A  is given by:
\begin{equation}
\label{eq19}
e_{\text{a}}[f] =
\begin{cases}
    K_{\text{m}} V_{\text{dc}} G_{\text{del}}(s) c_{\text{a}}[\pm f_{\text{p}}], & f = \pm f_{\text{p}} \\
    K_{\text{m}} V_{\text{dc}} G_{\text{del}}(s - 2j\omega_1) c_{\text{a}}[\pm f_{\text{p1}}], & f = \pm f_{\text{p1}}
\end{cases}
\end{equation}
where \(K_{\text{m}}\) is set to 0.5; $V_\mathrm{dc}$ denotes the dc-side voltage; \(G_{\mathrm{del}}(s)\) denotes the delay function, defined as \(G_{\mathrm{del}}(s)\) = \(e^{-1.5sT_{\mathrm{d}}}\), \(T_{\text{d}}\) denotes the sampling period, \(T_{\text{d}}\) = 1.5/\(f_\text{sw}\), \(f_\text{sw}\) is the switching frequency.

Substituting Eq. (\ref{eq19}) into Eq. (\ref{eq1}), the following expression can be obtained:
\begin{equation}
\label{eq20}
\begin{aligned}
    \begin{bmatrix}
        sL_{\text{f}} \mathbf{I}_{\text{p}} \\
        (s - 2j\omega_1) L_{\text{f}} \mathbf{I}_{\text{p1}}
    \end{bmatrix}
    &=
    \begin{bmatrix}
        K_{\text{m}} V_{\text{dc}} G_{\text{del}}(s) c_{\text{a}}[\pm f_{\text{p}}] \\
        K_{\text{m}} V_{\text{dc}} G_{\text{del}}(s - 2j\omega_1) c_{\text{a}}[\pm f_{\text{p1}}]
    \end{bmatrix} \\
    &\quad 
    -\begin{bmatrix}
        Y_{\text{con}}(s) \mathbf{V}_{\text{p}} \\
        Y_{\text{con}}(s - 2j\omega_1) \mathbf{V}_{\text{p1}}
    \end{bmatrix}
\end{aligned}
\end{equation}
where \(Y_{\text{con}}\)\((s\))=\(1+s^2 L_{\text{f}} C_{\text{f}}\).

Combining Eqs. (\ref{eq1}), (\ref{eq4}), and (\ref{eq20}), the impedance matrix can be derived as follows:
\begin{equation}
\label{eq21}
\begin{bmatrix}
    \mathbf{V}_{\text{p}} \\
    \mathbf{V}_{\text{p1}}
\end{bmatrix}
=
-
\begin{bmatrix}
    Z_{\text{inv}}^{\text{pp}} & Z_{\text{inv}}^{\text{pn}} \\
    Z_{\text{inv}}^{\text{np}} & Z_{\text{inv}}^{\text{nn}}
\end{bmatrix}
\begin{bmatrix}
    \mathbf{I}_{\text{p}} \\
    \mathbf{I}_{\text{p1}}
\end{bmatrix}
\end{equation}
where  \(Z_{\text{inv}}^{\text{pp}}\), \(Z_{\text{inv}}^{\text{pn}}\), \(Z_{\text{inv}}^{\text{np}}\), \(Z_{\text{inv}}^{\text{nn}}\) denote the corresponding elements of the DCI impedance matrix. The detailed expression is provided in the downloadable supplementary file \cite{website} [see Eqn. (A10) in the file].

\subsection{SISO Impedance Model for the DCI and Grid Side}
Under unbalanced conditions, the grid impedance matrix \( Z_{\text{g}} \)(\(s\)) and the grid-side load impedance matrix \( Z_{\text{L}} \)(\(s\)) are derived according to the symmetrical component method

\begin{equation}
\label{eq22}
Z_k(s) = \frac{1}{3}
\begin{bmatrix}
    Z_{ka} + Z_{kb} + Z_{kc} & Z_{ka} + a^2 Z_{kb} + a Z_{kc} \\
    Z_{ka} + a Z_{kb} + a^2 Z_{kc} & Z_{ka} + Z_{kb} + Z_{kc}
\end{bmatrix}
\qquad
\end{equation}
where \(a\) = \(e^{j 120^\circ}\); \(k\) = \(\text{g}\) represents the grid impedance matrix, i.e., \(Z_{\text{g}}\)(\(s\)); \(k\) = \(\text{L}\) represents the grid-side load impedance matrix, i.e., \(Z_{\text{L}}\)(\(s\)); \(Z_{k\text{a}}\), \(Z_{k\text{b}}\), \(Z_{k\text{c}}\) correspond to the three-phase impedance of the grid or loads. 

In power systems, three typical unbalanced operating conditions are encountered: (i) asymmetric loads; (ii) asymmetric grid impedance; and (iii) asymmetric grid voltages. According to the symmetrical component method and Eq. (\ref{eq22}), in case (i), \(Z_\text{L}\)(\(s\)) is a full matrix, while \(Z_\text{g}\)(\(s\)) is diagonal. In case (ii), \(Z_\text{g}\)(\(s\)) is a full matrix and \(Z_\text{L}\)(\(s\)) is diagonal. In case (iii), both matrices are diagonal.

The \( Z_{\text{g}}\)(\(s\)) and \( Z_{\text{L}} \)(\(s\)) are considered as a unified system, denoted as \( Z_{\text{grid}}\)(\(s\)), referred to as the grid-side impedance matrix. Since the grid and load impedances are connected in parallel, the equivalent impedance can be expressed as follows:
\begin{equation}
\label{eq23}
Z_{\text{grid}}(s) = Z_{\text{g}}(s) \parallel Z_{\text{L}}(s) =
\begin{bmatrix}
    Z_{\text{grid}}^{\text{pp}} & Z_{\text{grid}}^{\text{pn}} \\[4pt]
    Z_{\text{grid}}^{\text{np}} & Z_{\text{grid}}^{\text{nn}}
\end{bmatrix}
\end{equation}
where \(Z_{\text{grid}}^{\text{pp}}\), \(Z_{\text{grid}}^{\text{pn}}\), \(Z_{\text{grid}}^{\text{np}}\), \(Z_{\text{grid}}^{\text{nn}}\) are the corresponding elements of \(Z_{\text{grid}}\)(\(s\)).

To reduce the complexity of stability analysis, the multi-input multi-output models of the DCI and grid side are transformed into SISO models, with the coupling terms embedded into the equivalent positive- and negative-sequence circuits in the form of controlled sources \cite{zhang2017sequence}. The SISO impedances of the DCI and grid side can be expressed as follows:
\begin{equation}
\label{eq24}
Z_{\text{inv,p}} = Z_{\text{inv}}^{\text{pp}} - 
\frac{Z_{\text{inv}}^{\text{pn}} ( Z_{\text{inv}}^{\text{np}} + Z_{\text{grid}}^{\text{np}} )}
{Z_{\text{grid}}^{\text{nn}} + Z_{\text{inv}}^{\text{nn}}}
\end{equation}

\begin{equation}
\label{eq25}
Z_{\text{inv,n}} = Z_{\text{inv}}^{\text{nn}} - 
\frac{Z_{\text{inv}}^{\text{np}} ( Z_{\text{inv}}^{\text{pn}} + Z_{\text{grid}}^{\text{pn}} )}
{Z_{\text{grid}}^{\text{pp}} + Z_{\text{inv}}^{\text{pp}}}
\end{equation}

\begin{equation}
\label{eq26}
Z_{\text{grid,p}} = Z_{\text{grid}}^{\text{pp}} - 
\frac{Z_{\text{grid}}^{\text{pn}} ( Z_{\text{inv}}^{\text{np}} + Z_{\text{grid}}^{\text{np}} )}
{Z_{\text{grid}}^{\text{nn}} + Z_{\text{inv}}^{\text{nn}}}
\end{equation}

\begin{equation}
\label{eq27}
Z_{\text{grid,n}} = Z_{\text{grid}}^{\text{nn}} - 
\frac{Z_{\text{grid}}^{\text{np}} ( Z_{\text{inv}}^{\text{pn}} + Z_{\text{grid}}^{\text{pn}} )}
{Z_{\text{grid}}^{\text{pp}} + Z_{\text{inv}}^{\text{pp}}}
\end{equation}
where \(Z_{\text{inv,p}}\), \(Z_{\text{inv,n}}\), \(Z_{\text{grid,p}}\), and \(Z_{\text{grid,n}}\) represent the equivalent positive- and negative-sequence impedances of the DCI and the grid side, respectively.

\begin{table}[!t]
\caption{Parameters of the grid-connected system\label{tab:table1}}
\centering
\renewcommand{\arraystretch}{1} 
\setlength{\tabcolsep}{3pt}  
\begin{tabular}{p{2.85cm} c||p{2.85cm} c}
\hline
\text{Parameters} & \text{Value} & \text{Parameters} & \text{Value} \\
\hline
$L_{\text{f}}$ (mH) & 3 & $k_{\text{pv}}$ & 2 \\
$C_{\text{f}}$ ($\mu$F) & 20 & $k_{\text{iv}}$ & 50 \\
$V_{\text{dc}}$ (V) & 700 & $k_{\text{pi}}$ & 2 \\
$P_{\text{ref}}$ (kW) & $10$ & $k_{\text{ii}}$ & 100 \\
$Q_{\text{ref}}$~(kvar) & 0 & $R_{\text{g}}$~($\Omega$) & 0.2 \\
$m$ & $5 \times 10^{-4}$ & $L_{\text{g}}$ (mH) & 3 \\
$n$ & $5 \times 10^{-4}$ & $V_{\text{n}}$ (V) & 311 \\
$f_{\text{sw}}$ (Hz) & $10^{4}$ & $V_2$ (V) & 30.9 \\
$T_{\text{d}}$ & $1.5/f_{\mathrm{sw}}$ & $\omega_{\text{n}}$ & $100\pi$ \\
\hline
\end{tabular}
\end{table} 
\subsection{Validation of Impedance Model Equivalence}
The parameters of the grid-connected system are listed in Table \MakeUppercase{\romannumeral 1}. To verify the accuracy of the established impedance model, an impedance measurement simulation platform is built in MATLAB/Simulink. Subsequently, three-phase positive- and negative-sequence perturbation voltages are injected separately at the PCC in two steps to measure the impedance of the DCI and the grid side, respectively \cite{elshenawy2024analysis}.

The signal-to-noise ratio (SNR) is a critical factor affecting the accuracy of impedance measurements in MATLAB/Simulink. To ensure reliable measurement, a frequency-sweeping approach based on single-sinusoidal disturbance injection is adopted, which have been proven to achieve a high SNR \cite{roinila2018hardware}. Moreover, the impedances of the DCI and the grid side are measured three times and averaged to reduce random errors. Finally, the method in \cite{nian2021design} is utilized to determine the minimum SNR requirement based on the maximum allowable measurement error, providing a quantitative guideline for designing the disturbance amplitude. The allowable magnitude and phase errors are set to 2 dB and \(8^\circ\), respectively.

The impedance measurement results for the DCI and grid side are presented in Figs. 2 and 3, with the grid voltage unbalance set at 10\%. As illustrated in Figs. 2 and 3, the red solid lines denote the theoretical equivalent positive-sequence impedance, while the red circles indicate the corresponding measurement results. Similarly, the blue solid lines represent the theoretical equivalent negative-sequence impedance, and the blue circles represent the corresponding measurement results. Overall, the measurement results are in good agreement with the theoretical models. 

Further analysis reveals that the equivalent positive-sequence impedance of the DCI exhibits capacitive negative resistance in the sub/super-synchronous frequency ranges (0--100Hz), whereas the grid impedance remains inductive (with a phase angle between 0° and 90°). This behavior may induce interactive oscillations between the DCI and the grid, potentially leading to oscillatory risks. In contrast, the equivalent negative-sequence impedance of the DCI does not exhibit the capacitive negative resistance, indicating relatively stable behavior.

\begin{figure}
    \centering
    \includegraphics[width=1\linewidth]{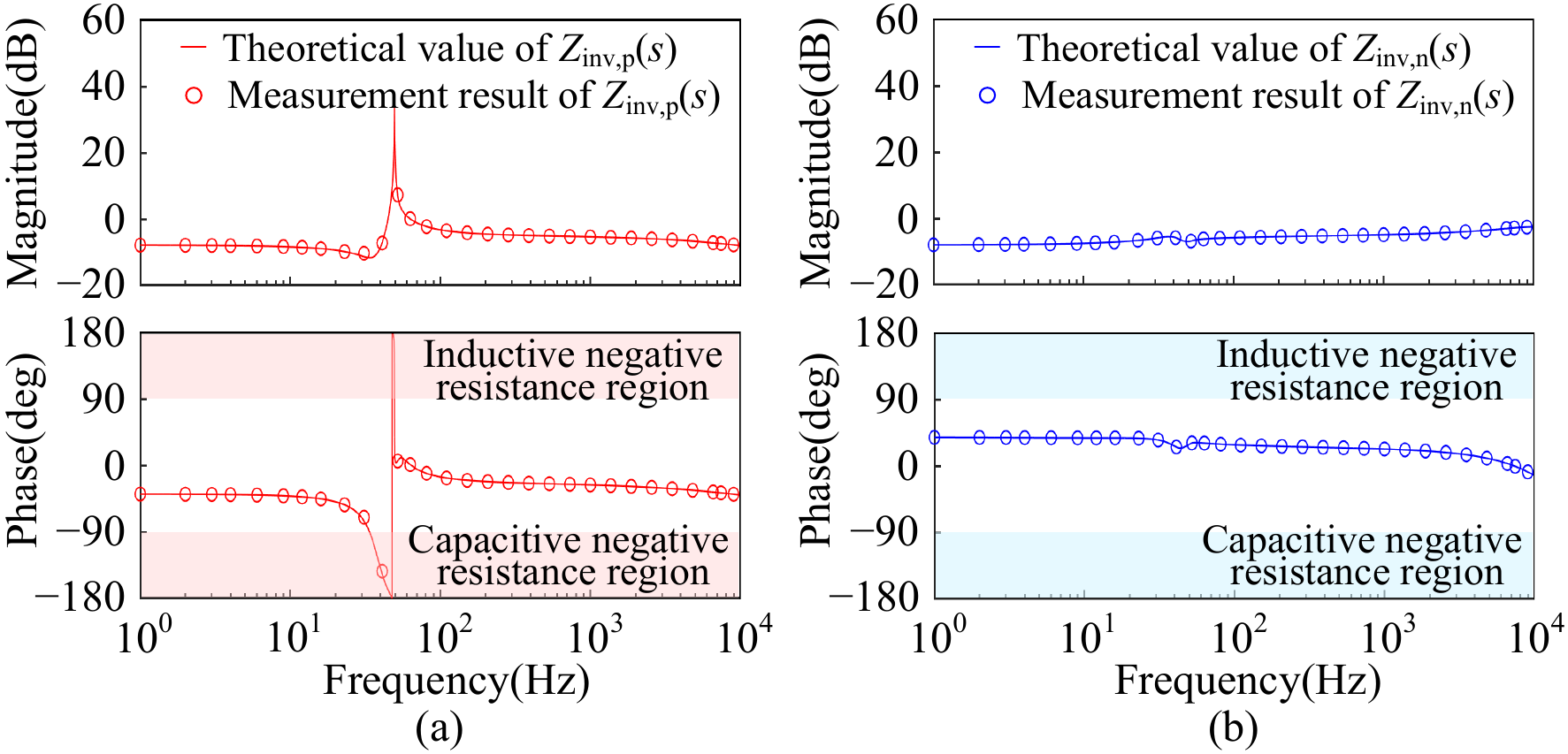}
    \caption{Established SISO impedance of the DCI and simulation-based impedance measurement results. (a) Established equivalent positive-sequence impedance of the DCI and simulation-based impedance measurement results. (b) Established equivalent negative-sequence impedance of the DCI and simulation-based impedance measurement results.}
    \label{fig.2}
\end{figure}

\begin{figure}
    \centering
    \includegraphics[width=1\linewidth]{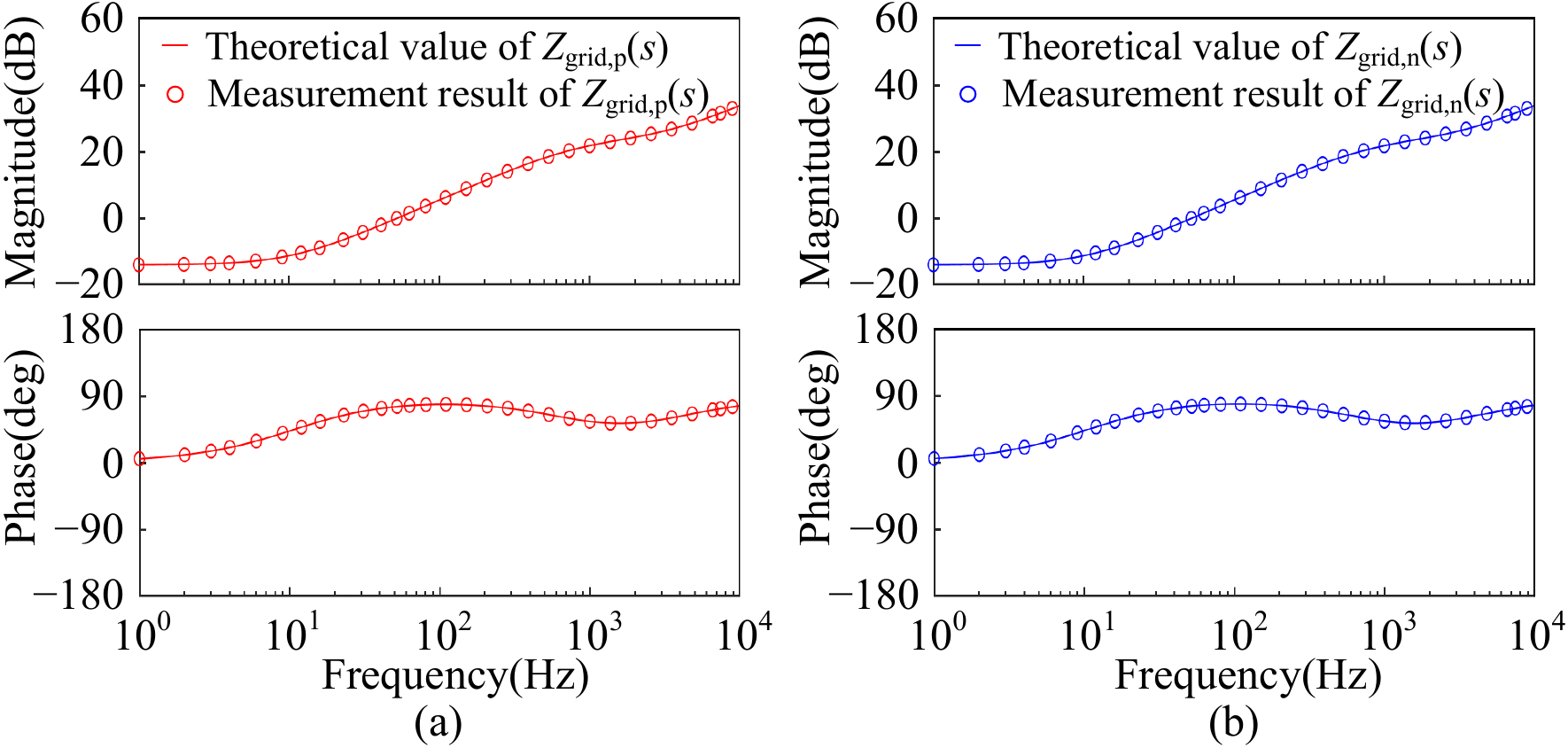}
    \caption{Established SISO impedance of the grid side and simulation-based impedance measurement results. (a) Established equivalent positive-sequence impedance of the grid side and simulation-based impedance measurement results. (b) Established equivalent negative-sequence impedance of the grid side and simulation-based impedance measurement results.}
    \label{fig.3}
\end{figure}

\subsection{Comparison with State-of-the-Art Models}
Fig. 4 compares the proposed model with those reported in \cite{zhao2024impedance,yan2021sequence,elshenawy2023unified} and \cite{rong2023modeling,elshenawy2024analysis,hu2019sequence}. The red circles represent the measured impedance values of the DCI or grid side. The proposed model incorporates both MFCE and the fundamental negative-sequence component, whereas the models in \cite{zhao2024impedance,yan2021sequence,elshenawy2023unified} consider neither, and those in \cite{rong2023modeling,elshenawy2024analysis,hu2019sequence} include only MFCE. The simulation results are obtained under unbalanced grid impedance conditions, with \(L_{\text{ga}}\) = \(L_{\text{gb}}\) = 3 \(\text{mH}\) and \(L_{\text{gc}}\) = 10.5 \(\text{mH}\), while the remaining parameters are consistent with Table I. 

\begin{figure}
    \centering
    \includegraphics[width=1\linewidth]{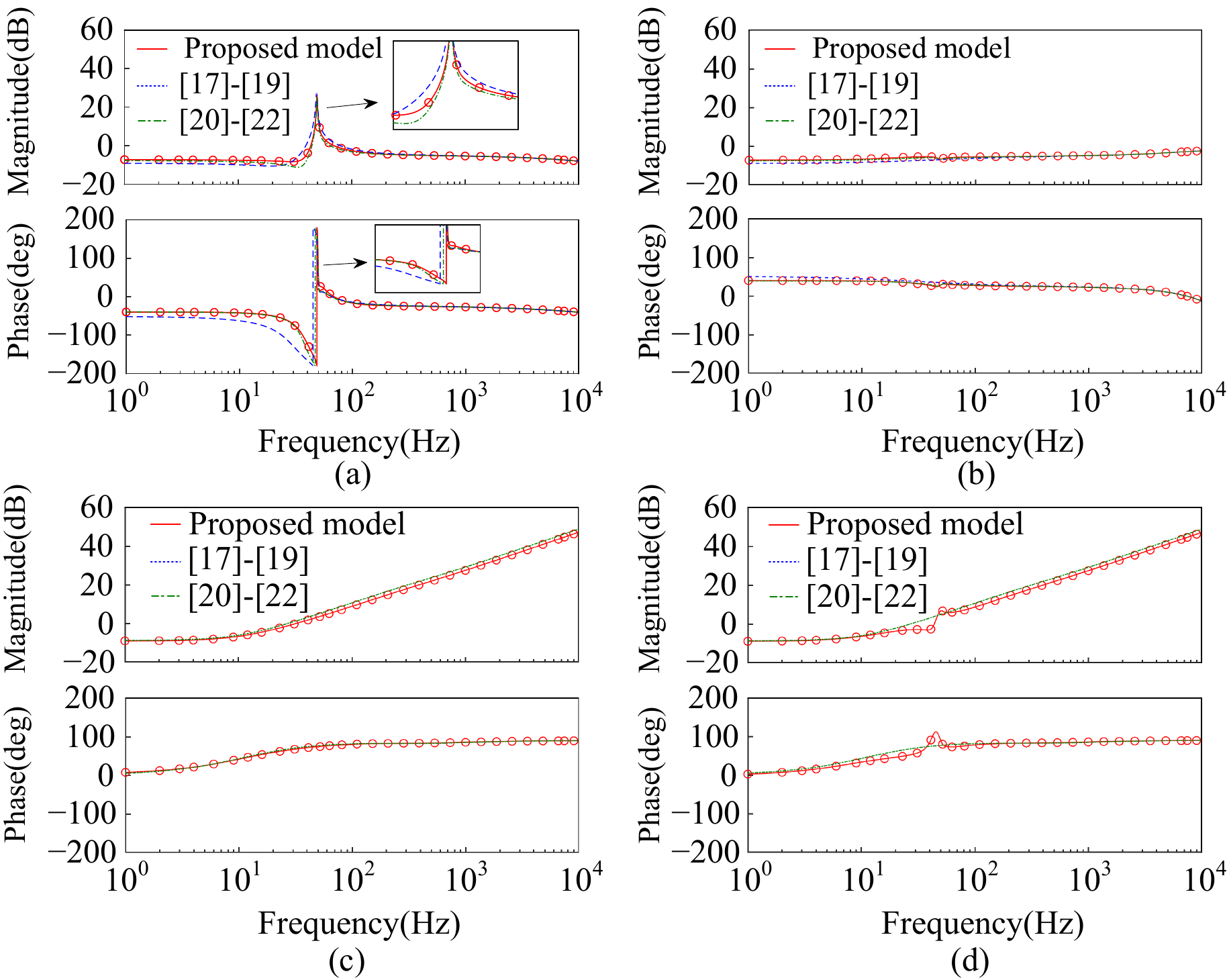}
   \caption{Comparison between the proposed model and the models in \cite{zhao2024impedance,yan2021sequence,elshenawy2023unified} and \cite{rong2023modeling,elshenawy2024analysis,hu2019sequence}. (a) Bode plot of the equivalent positive-sequence impedance of the DCI. (b) Bode plot of the equivalent negative-sequence impedance of the DCI. (c) Bode plot of the equivalent positive-sequence impedance of the grid side. (c) Bode plot of the equivalent negative-sequence impedance of the grid side. }
    \label{fig.4}
\end{figure}

A comparison between the models in \cite{zhao2024impedance,yan2021sequence,elshenawy2023unified} and \cite{rong2023modeling,elshenawy2024analysis,hu2019sequence} reveals that MFCE primarily affects the DCI impedance within the 1--100 Hz range, whereas its impact on the grid-side impedance is negligible, as the blue and green dashed curves in Figs. 4(c) and 4(d) nearly overlap. Further comparison between \cite{rong2023modeling,elshenawy2024analysis,hu2019sequence} and the proposed model indicates that the fundamental negative-sequence component has a dominant influence on the DCI impedance in the 1--50 Hz range under unbalanced grid impedance conditions, as shown in Figs. 4(a) and 4(b). Moreover, since the models in \cite{zhao2024impedance,yan2021sequence,elshenawy2023unified} and \cite{rong2023modeling,elshenawy2024analysis,hu2019sequence} neglect the coupling between positive- and negative-sequence components, they fail to accurately represent the grid-side impedance characteristics under unbalanced conditions, as illustrated in Figs. 4(c) and 4(d). Therefore, to model both DCI and grid-side impedance under unbalanced grid impedance conditions accurately, it is essential to consider both MFCE and the fundamental negative-sequence component.

\subsection{Impact of Grid Voltage Unbalance Level on MFCE and Impedance Characteristics}
Based on the developed impedance models, this section investigates the impact of grid voltage unbalance level on MFCE and the impedance characteristics. The grid voltage unbalanced level \(\delta\) is defined in Eq. (\ref{eq28}). To evaluate the effects of different unbalanced conditions, five sets of fundamental negative-sequence voltages are applied, corresponding to unbalance levels of 5\%, 10\%, 15\%, 20\%, and 25\%. During the simulations, the three-phase loads and grid impedance are maintained symmetric, while the settings of other parameters are listed in Table~I.
\begin{equation}
\label{eq28}
\delta= \frac{V_2}{V_1} \times 100\% \
\end{equation}
where \(V_1\) is 311 V.

To intuitively illustrate the impact of the \( \delta \) on the MFCE, a comprehensive frequency coupling coefficient \( \varepsilon \) is introduced as a quantitative indicator, as defined in Eq. (\ref{eq29}). Based on the physical meanings of $Z_{\text{inv}}^{\text{pn}}$ and $Z_{\text{inv}}^{\text{pp}}$ \cite{zhang2017sequence}, their ratio reflects the magnitude of the coupling current response $\textbf{I}_\text{p1}$ relative to the perturbation current response $\textbf{I}_\text{p}$. Therefore, the magnitude of \( \varepsilon \) can be used to quantify the strength of the MFCE. A larger frequency-domain magnitude of \( \varepsilon \) indicates a stronger MFCE. As illustrated in Fig. 5, the magnitude of \( \varepsilon \) attains its maximum in the vicinity of \(f_1\), while remaining relatively small across other frequency ranges. This suggests that the fundamental negative-sequence component has a dominant influence on the MFCE in the vicinity of \(f_1\), which may affect the local impedance characteristics around the fundamental frequency.
\begin{equation}
\label{eq29}
\varepsilon = 20 \log_{10} \frac{\left| Z_{\text{inv}}^{\text{pn}} \right|}{\left| Z_{\text{inv}}^{\text{pp}} \right|}
\end{equation}

Fig. 6 illustrates the Bode plots of the equivalent positive- and negative-sequence impedance of the DCI under different grid voltage unbalance levels. As \(\delta\) increases, the resonance peak of the equivalent positive-sequence impedance near \(f_1\) diminishes, whereas the impact on the phase angle remains minor. Additionally, the magnitude and phase of the equivalent negative-sequence impedance remain nearly unchanged. These results indicate that grid voltage unbalance primarily affects the magnitude of the equivalent positive-sequence impedance of the DCI near \(f_1\), with little influence on its phase angle or the equivalent negative-sequence impedance.

\begin{figure}
    \centering
    \includegraphics[width=1\linewidth]{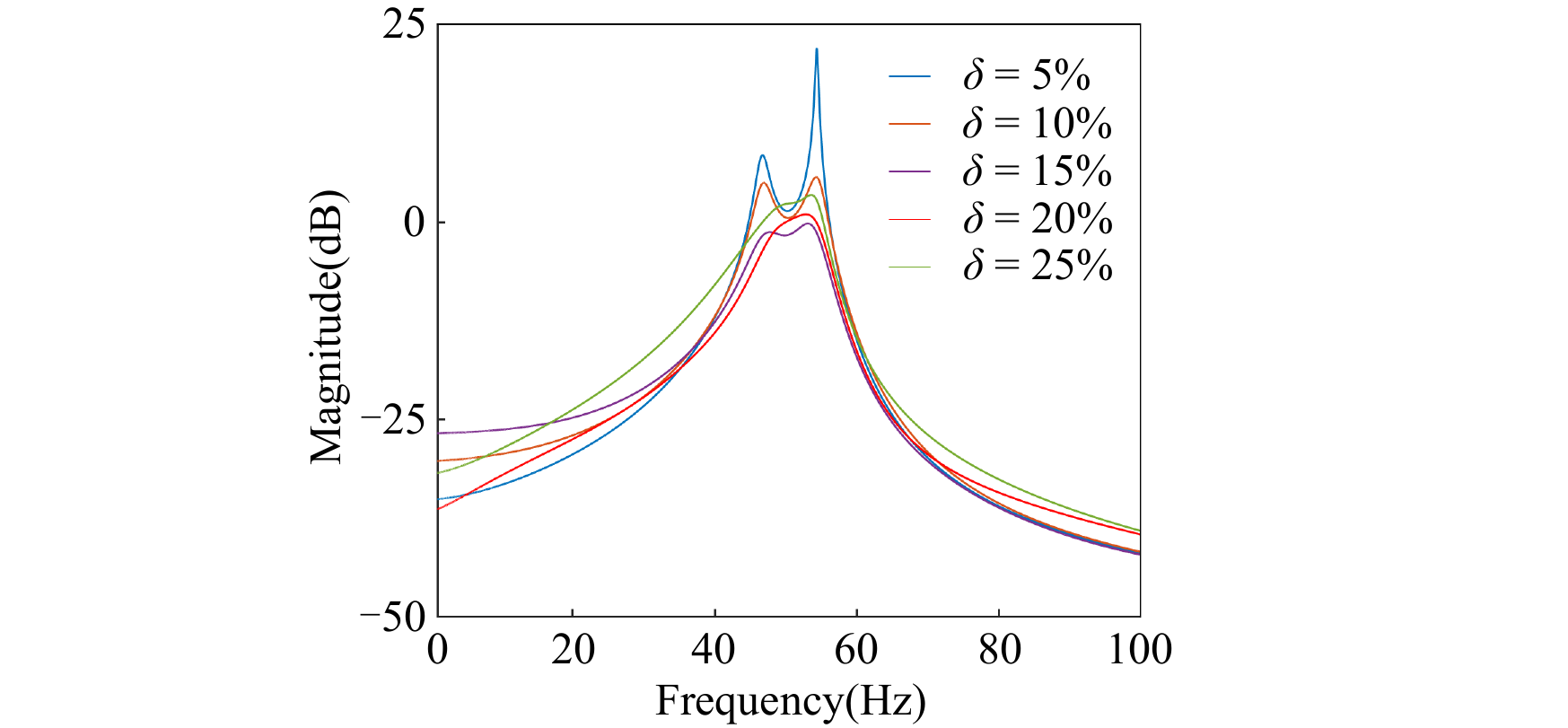}
   \caption{Magnitude curves of $\varepsilon$ with different grid voltage unbalance levels.}
    \label{fig.5}
\end{figure}
\begin{figure}
    \centering
    \includegraphics[width=1\linewidth]{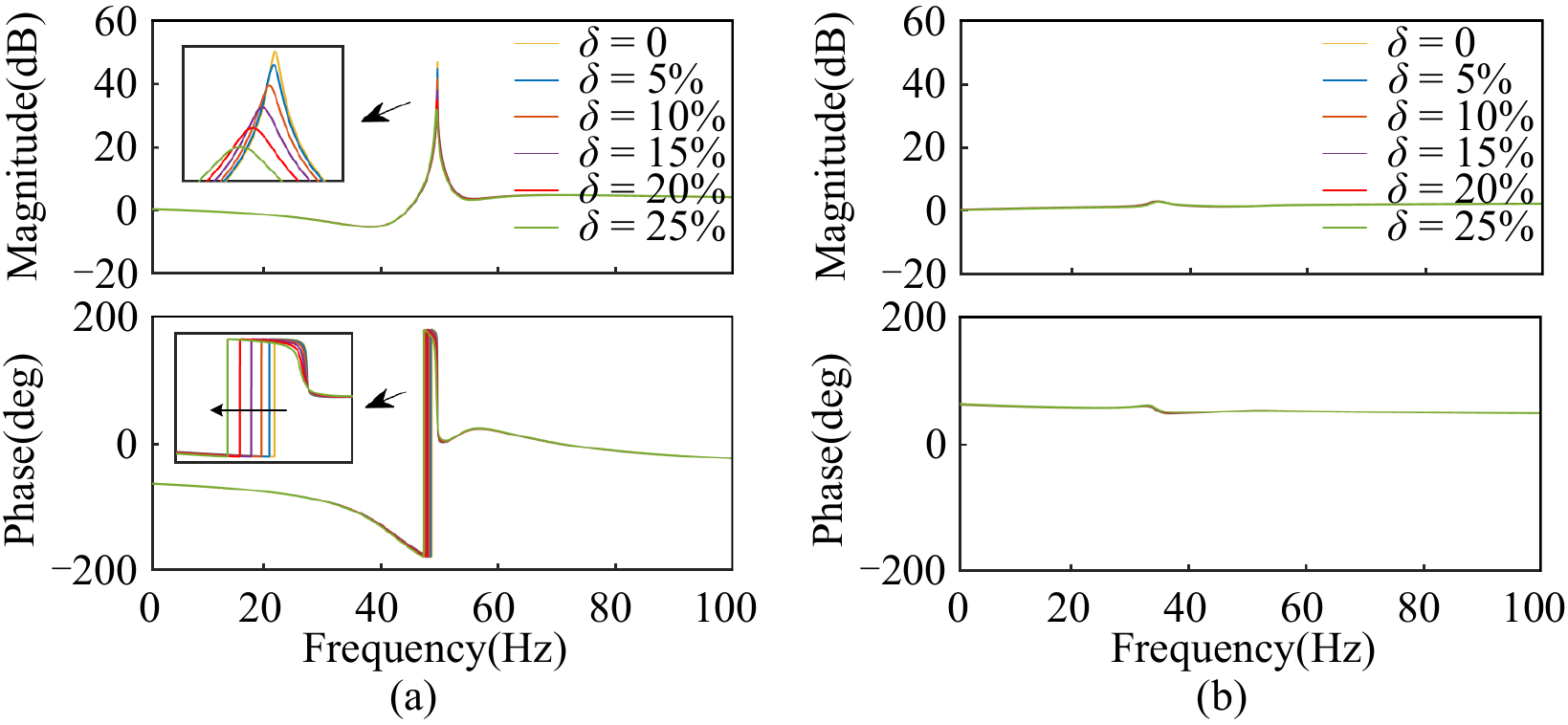}
   \caption{Equivalent positive and negative-sequence impedance Bode diagrams of the DCI under different grid voltage unbalanced levels. (a) Equivalent positive-sequence impedance Bode diagram of the DCI. (b) Equivalent negative-sequence impedance Bode diagram of the DCI.}
    \label{fig.6}
\end{figure}

\section{Stability Analysis for Grid-Connected Systems Under Unbalanced Conditions}
\subsection{Identification of Dominant Stability Factors}
Based on the impedance models established in Section \textup{II} and Nyquist criterion \cite{li2022systematic, zhu2022impedance}, the following relationship can be obtained:
\begin{equation}
\label{eq30}
\lambda_{\text{p}} = \frac{Z_{\text{inv,p}}}{Z_{\text{grid,p}}}
\end{equation}
\begin{equation}
\label{eq31}
\lambda_{\text{n}} = \frac{Z_{\text{inv,n}}}{Z_{\text{grid,n}}}
\end{equation}
where \(\lambda_{\text{p}}\) and \(\lambda_{\text{n}}\) are the positive- and negative-sequence impedance ratio, respectively.

Based on the impedance model of the DCI, $\lambda_{\text{p}}$ and $\lambda_{\text{n}}$ are mainly affected by circuit parameters, control parameters, and the grid impedance. An improved normalized sensitivity analysis method is proposed to quantitatively evaluate the influence of key parameters on the stability of grid-connected systems. Specifically, the influence of parameter $\alpha$ on system stability is evaluated based on its average normalized sensitivity over a small variation interval. The normalized sensitivity is defined as follows:
\begin{equation}
\label{eq32}
L_x(\alpha, s) 
= \frac{\left[\lambda_x\!\bigl(\alpha + \Delta\alpha, s\bigr) 
- \lambda_x(\alpha, s)\right]/\Delta\alpha}%
{\lambda_x(\alpha, s)/\alpha}
\end{equation}
where \(L_{x}\)(\(\alpha, s\)) denotes the normalized sensitivity of the positive- or negative-sequence impedance ratio; \(x\) = p and \(x\) = n represent the positive and negative-sequence, respectively; \(\lambda_{x}\)(\(\alpha, s\)) denotes the positive- or negative-sequence impedance ratio; $\Delta\alpha$ denotes the minor change in $\alpha$, $\Delta\alpha$ = [$-b$, $b$]$\cdot\alpha$, the value of $b$ is set to 10$\%$.

Eq. (\ref{eq32}) can be decomposed into its real and imaginary components, corresponding to the resistance sensitivity \( L_{x,\text{Re}}\)(\(\alpha, s\)) and the reactance sensitivity \( L_{x,\text{Im}}\)(\(\alpha, s\)).
\begin{equation}
\label{eq33}
\begin{cases}
L_{x,\text{Re}}(\alpha, s) = \operatorname{Re} \{ L_x(\alpha, s) \} \\
L_{x,\text{Im}}(\alpha, s) = \operatorname{Im} \{ L_x(\alpha, s) \}
\end{cases}
\end{equation}

To quantify the impact of different parameters, a weighting analysis method is introduced to calculate the proportion of resistance sensitivity, denoted as \( L_{x,\text{Re}\%}\)(\(\alpha\)), and the proportion of reactance sensitivity, denoted as \( L_{x,\text{Im}\%}\)(\(\alpha\)). The corresponding calculation method is given as follows:
\begin{equation}
\label{eq34}
\begin{cases}
L_{x,\text{Re}\%}(\alpha) = \dfrac{\bar{L}_{x,\text{Re}}(\alpha)}{\displaystyle \sum \bar{L}_{x,\text{Re}}} \times 100\% \\[14pt]
L_{x,\text{Im}\%}(\alpha) = \dfrac{\bar{L}_{x,\text{Im}}(\alpha)}{\displaystyle \sum \bar{L}_{x,\text{Im}}} \times 100\%
\end{cases}
\end{equation}
where \( \bar{L}_{x,\text{Re}}\)(\(\alpha\)) and \( \bar{L}_{x,\text{Im}}\)(\(\alpha\)) represent the averaged normalized sensitivities of resistance and reactance with respect to parameter \( \alpha \), respectively. The sensitivities are evaluated at a set of equally spaced points within the variation interval $\Delta\alpha$ and averaged (additional details are provided in Section S6-A of the supplementary material \cite{website}). The terms \( \sum \bar{L}_{x,\text{Re}} \) and \( \sum \bar{L}_{x,\text{Im}} \) denote the total sum of the averaged normalized sensitivities of resistance and reactance for all parameters, respectively.
\begin{figure}
    \centering
    \includegraphics[width=1\linewidth]{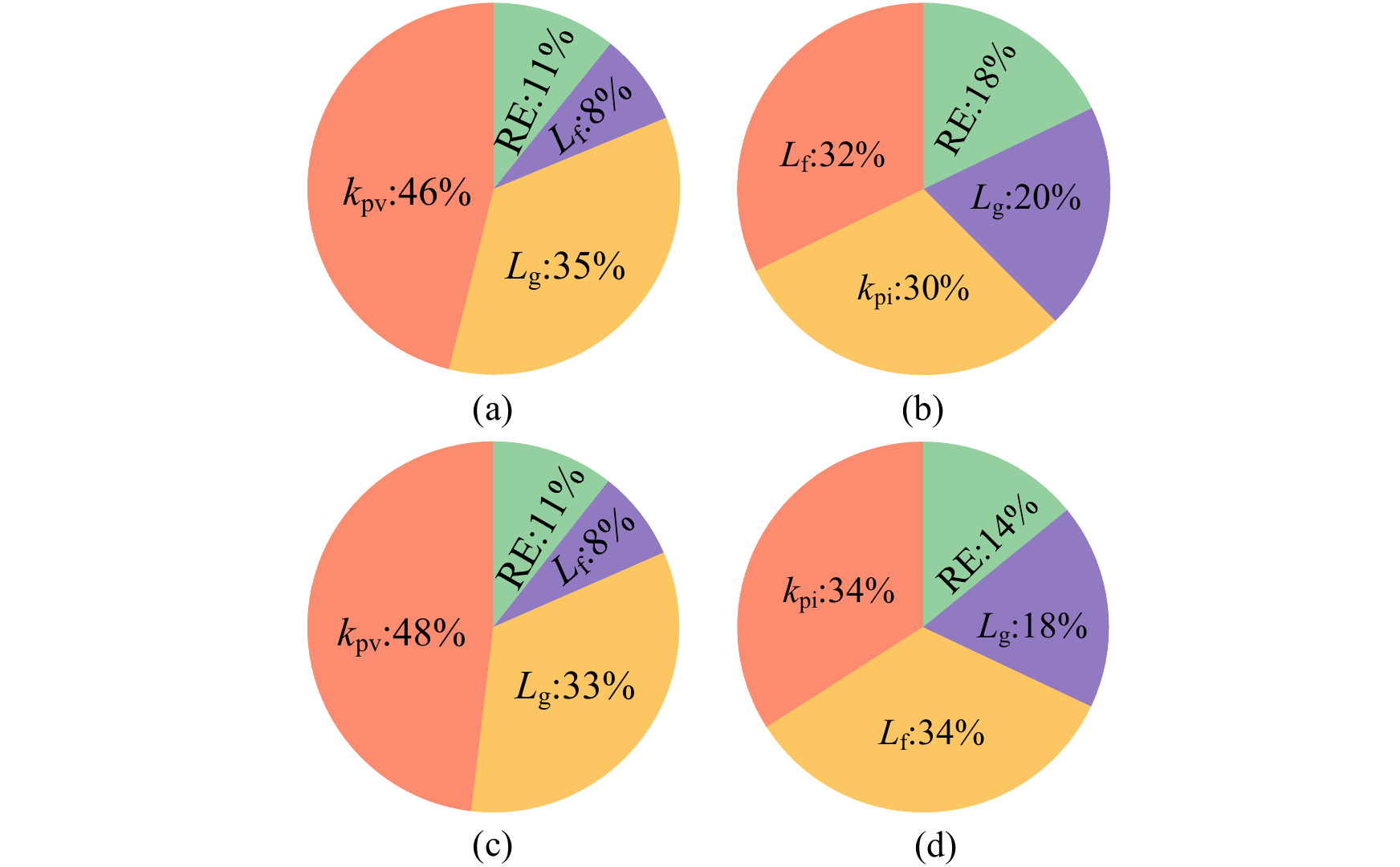}
   \caption{Proportional chart of resistance/reactance normalized sensitivity for $\lambda_{\text{p}}$ and $\lambda_{\text{n}}$. (a) Normalized sensitivity distribution of resistance for $\lambda_{\text{p}}$. (b) Normalized sensitivity distribution of reactance for $\lambda_{\text{p}}$. (c) Normalized sensitivity distribution of resistance for $\lambda_{\text{n}}$. (d) Normalized sensitivity distribution of reactance for $\lambda_{\text{n}}$.}
    \label{fig.7}
\end{figure}

According to Eq. (\ref{eq34}), the proportion of resistance and reactance sensitivity is computed for each parameter, i.e., $m$, $n$, $k_{\text{pv}}$, $k_{\text{iv}}$, $k_{\text{pi}}$, $k_{\text{ii}}$, $L_{\text{g}}$, $V_2$, $L_{\text{f}}$, $C_{\text{f}}$, and $T_{\text{d}}$. The corresponding parameter values are provided in Table I. Fig. 7 depicts the top three parameters with the highest normalized sensitivity ratios for resistance and reactance. The remaining parameters, which exhibit relatively minor contributions, are categorized under ``RE". The results indicate that system stability is predominantly affected by $k_{\text{pv}}$, $k_{\text{pi}}$, $L_{\text{g}}$, and $L_{\text{f}}$ within the sub/super-synchronous frequency ranges, as shown in Fig. 7. In contrast, other parameters exert comparatively less influence on system stability. It should be noted that the sensitivity  analysis is performed around a fixed parameter operating point and parameter set. If the system's operating conditions or parameters change, the dominant parameters may also vary (additional details are provided in Section S6-B of the supplementary material \cite{website}).
\subsection{System Stability Analysis}
The Nyquist and Bode criteria are widely applied to evaluate stability. Since Nyquist curves are polar plots of eigenvalues, the frequency information can not be obtained directly from the Nyquist curves\cite{yang2020stability}. In contrast, the Bode plot provides a clear representation of the system’s frequency-domain characteristics, including the frequency-dependent magnitude and phase information. Since neither $\lambda_{\text{p}}$ nor $\lambda_{\text{n}}$ has right-half-plane poles, the Bode criterion can be applied (additional details are provided in Section S3 in the file) \cite{website}. The Bode criterion evaluates system stability by separately examining the Bode plots of the DCI's impedance \(Z_{\text{inv}}\) and the grid-side impedance \(Z_{\text{grid}}\). The magnitude intersection of \(Z_{\text{inv}}\) and \(Z_{\text{grid}}\) predicts potential resonances, and phase margin (PM) is defined based on the phase difference at the magnitude intersection as given in (\ref{eq35}) \cite{10994337,liu2020sequence}.

\begin{equation}
\label{eq35}
\text{PM} = 180^\circ - \left| \angle Z_{\text{inv}}(j\omega) - \angle Z_{\text{grid}}(j\omega) \right|
\end{equation}
where \(\angle Z_{\text{inv}}\)(\(j\omega\)) and and \(\angle Z_{\text{grid}}\)(\(j\omega\)) denote the phase angles of the DCI and grid-side impedances at the magnitude intersection, respectively.

The stability of a grid-connected system primarily depends on the phase margins of the positive- and negative-sequence systems, denoted as $\text{PM}_\text{p}$ and $\text{PM}_\text{n}$, respectively. $\text{PM}_\text{p}$ is defined as the phase angle difference between the DCI and grid-side equivalent positive-sequence impedances at their magnitude intersection, while $\text{PM}_\text{n}$ is defined as the corresponding phase angle difference for the equivalent negative-sequence impedances. As shown in Figs. 2 and 3, the phase angles of the DCI and grid-side equivalent negative-sequence impedances remain within 0°, --90°, ensuring that $\text{PM}_\text{n}$ is always positive and thereby guaranteeing the stability of the negative-sequence system. Therefore, the overall system stability mainly relies on $\text{PM}_\text{p}$. In general, a negative $\text{PM}_\text{p}$ indicates that the system is unstable. However, the system may remain unstable even with a positive phase margin close to 0°, owing to the insufficient stability margin. In this paper, the system is considered to have adequate stability margin when $\text{PM}_\text{p} \ge 10$° \cite{yang2020stability}. The threshold of 10° is determined based on the minimum acceptable damping ratio of 0.05, with an additional margin for conservativeness \cite{10994337}.
 
\begin{figure}
    \centering
    \includegraphics[width=1\linewidth]{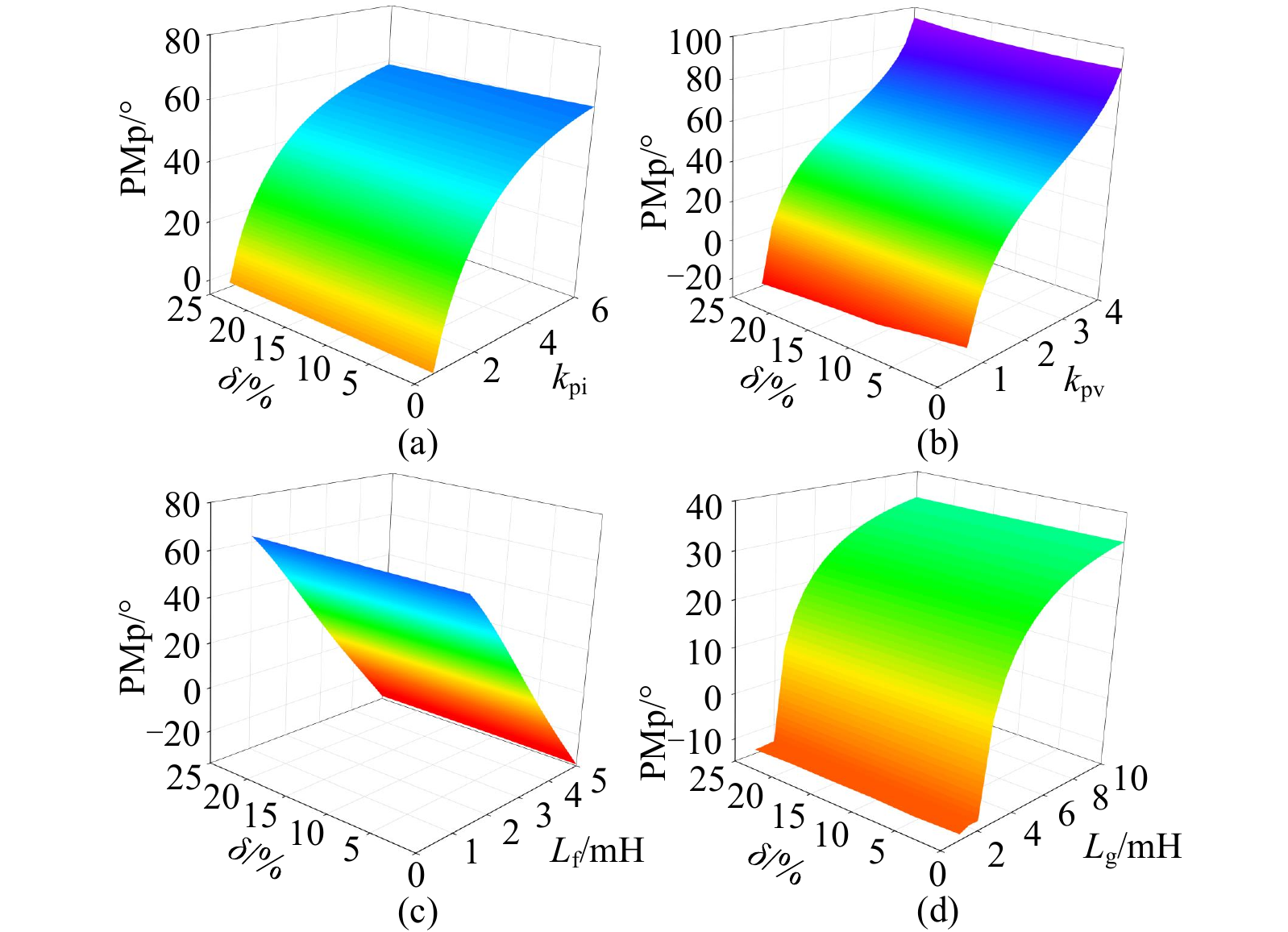}
   \caption{Effect of \(k_{\text{pi}}\), \(k_{\text{pv}}\), \(L_{\text{f}}\) and \(L_{\text{g}}\) variations on the phase margin of the positive-sequence system under different values of \(\delta\).}
    \label{fig.8}
\end{figure}

\begin{figure}
    \centering
    \includegraphics[width=1\linewidth]{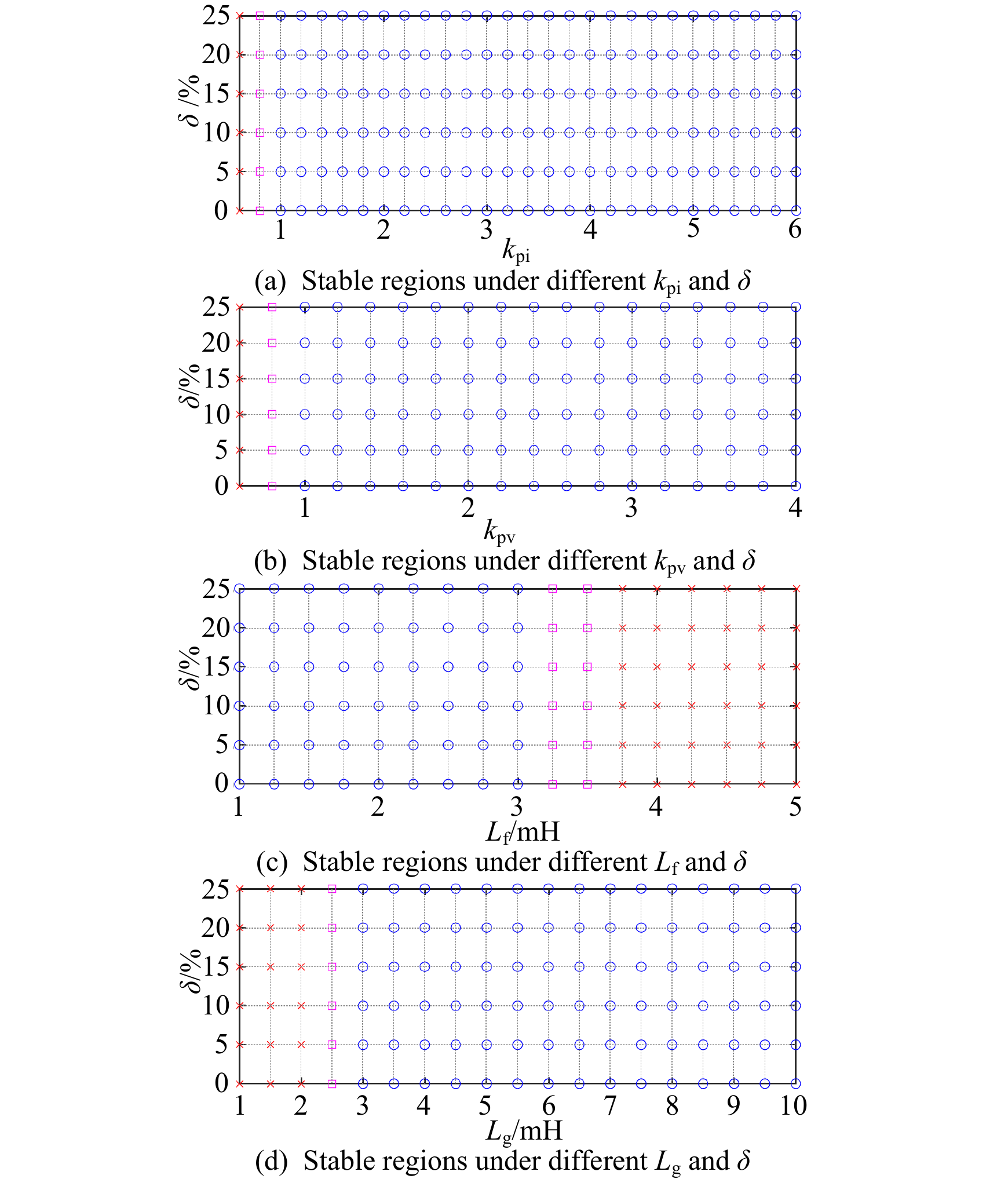}
   \caption{Stable regions under different parameters and \(\delta\).}
    \label{fig.9}
\end{figure}

\begin{figure}
    \centering
    \includegraphics[width=1\linewidth]{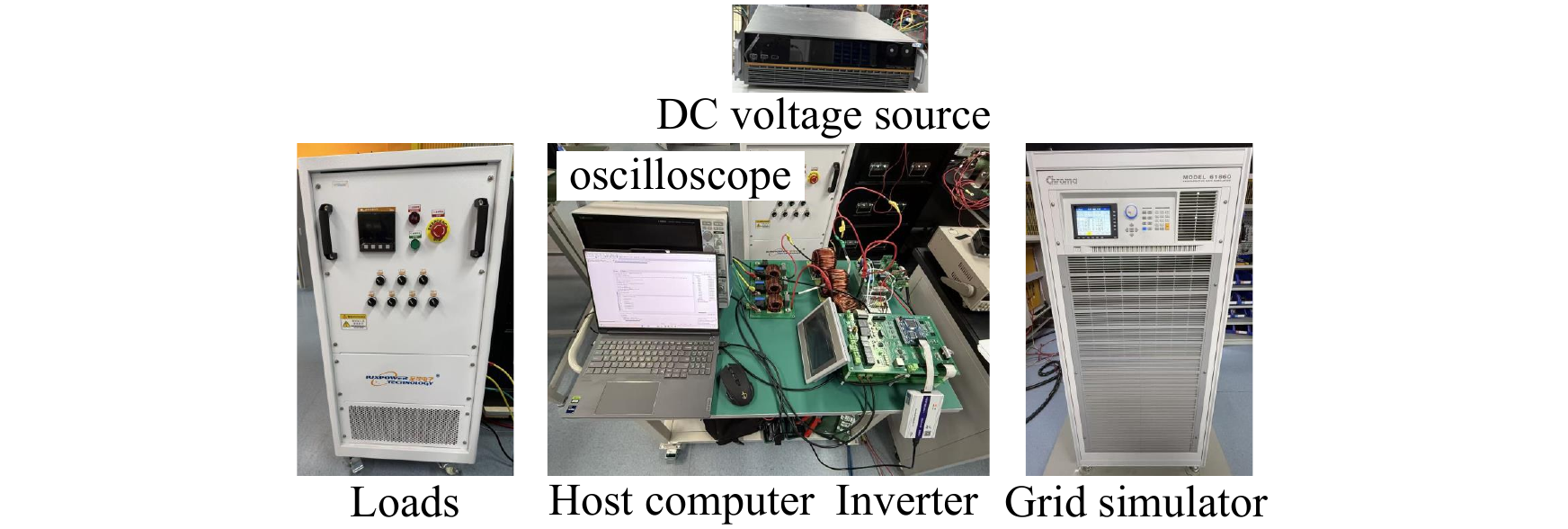}
   \caption{The experimental platform of droop-controlled grid-connected inverter.}
    \label{fig.10}
\end{figure}

This work considers grid voltage unbalance as a case study to evaluate the influence of \(k_{\text{pi}}\), \(k_{\text{pv}}\), \(L_{\text{f}}\), and \(L_{\text{g}}\) on system stability under varying degrees of grid voltage unbalance. According to Section III-A, these parameters exert a dominant influence on system stability. Additional analyses for unbalanced load and grid impedance conditions are provided in Section S4 of the supplementary material \cite{website}.

Fig. 8 illustrates the influence of variations in \(k_{\text{pi}}\), \(k_{\text{pv}}\), \(L_{\text{f}}\) and \(L_{\text{g}}\) on $\text{PM}_\text{p}$ under different levels of grid voltage unbalance. The parameters are varied as follows: \(k_{\text{pi}}\) from 0.6 to 6.0 in steps of 0.2, \(k_{\text{pv}}\) from 0.6 to 4.0 in steps of 0.2, \(L_{\text{f}}\) from 1 mH to 5 mH in steps of 0.25 mH, \(L_{\text{g}}\) from 1 mH to 10 mH in steps of 0.5 mH, and \(\delta\) from 0 to 25\% in steps of 5\%. As shown in Figs. 8(a) and 8(b), for a fixed \(\delta\), increasing \(k_{\text{pi}}\) or \(k_{\text{pv}}\) enhances the system stability margin. In contrast, when \(k_{\text{pi}}\) or \(k_{\text{pv}}\) is fixed, variations in \(\delta\) exert only a negligible effect. Similarly, according to Figs. 8(c) and 8(d), for a fixed \(\delta\), increasing \(L_{\text{g}}\) or decreasing \(L_{\text{f}}\) improves the system's stability margin. When either \(L_{\text{f}}\) or \(L_{\text{g}}\) is fixed, altering \(\delta\) has only a negligible effect on system margin. In summary, the results indicate that grid voltage unbalance has a negligible impact on system stability. This is mainly because different levels of unbalance have little effect on the phase angle of the DCI equivalent positive-sequence impedance, as shown in Fig. 6(a). 

To visualize the relationship between the four key parameters and the system stability margin under different values of \(\delta\), the corresponding stability regions are plotted in Fig. 9. The symbol ``\scalebox{0.7}{$\textcolor{blue}{\bigcirc}$}" denotes parameter combinations with sufficient stability margin ($\text{PM}_\text{p}$ \(\ge\)10°), which represents the recommended parameter combinations for stability; ``\textcolor{magenta}{\(\small\square\)}" denotes critically stable combinations (0° \(\le\) $\text{PM}_\text{p}$\(<\)10°), indicating insufficient stability margin; ``\textcolor{red}{\(\times\)}" represents unstable combinations ($\text{PM}_\text{p}$\(<\)0°). To ensure adequate phase margins, the proportional gains of current loop and voltage loop are selected to satisfy $k_\text{pi} \ge 1$ and $k_\text{pv} \ge 1$, while the filter inductance is designed to satisfy $ 1\le L_\text{f} \le 3~\mathrm{mH}$ and the grid inductance is required to meet $L_\text{g} \ge 3~\mathrm{mH}$, corresponding to a short-circuit ratio (SCR) \cite{elshenawy2023unified} of 2.5, indicating a weak grid condition. Notably, the threshold can be flexibly adjusted according to practical requirements. As the threshold increases, the critical stability region in Fig. 9 expands, the stable region shrinks accordingly, and the unstable region remains unchanged. In this case, although the range of the key parameters needs to be adjusted more strictly, the basic direction of parameter tuning follows the same trend.

\subsection{Experimental Validation}

To further verify the accuracy of the theoretical analysis, an experimental platform for the droop-controlled grid-connected inverter is constructed, as shown in Fig. 10. The control algorithm is implemented on the DSP TMS320F28335. The experimental platform consists mainly of a DC voltage source, three-phase loads, an inverter, a grid simulator, a host computer, and an oscilloscope.

\begin{figure}
	\centering\
	\includegraphics[width=1\linewidth]{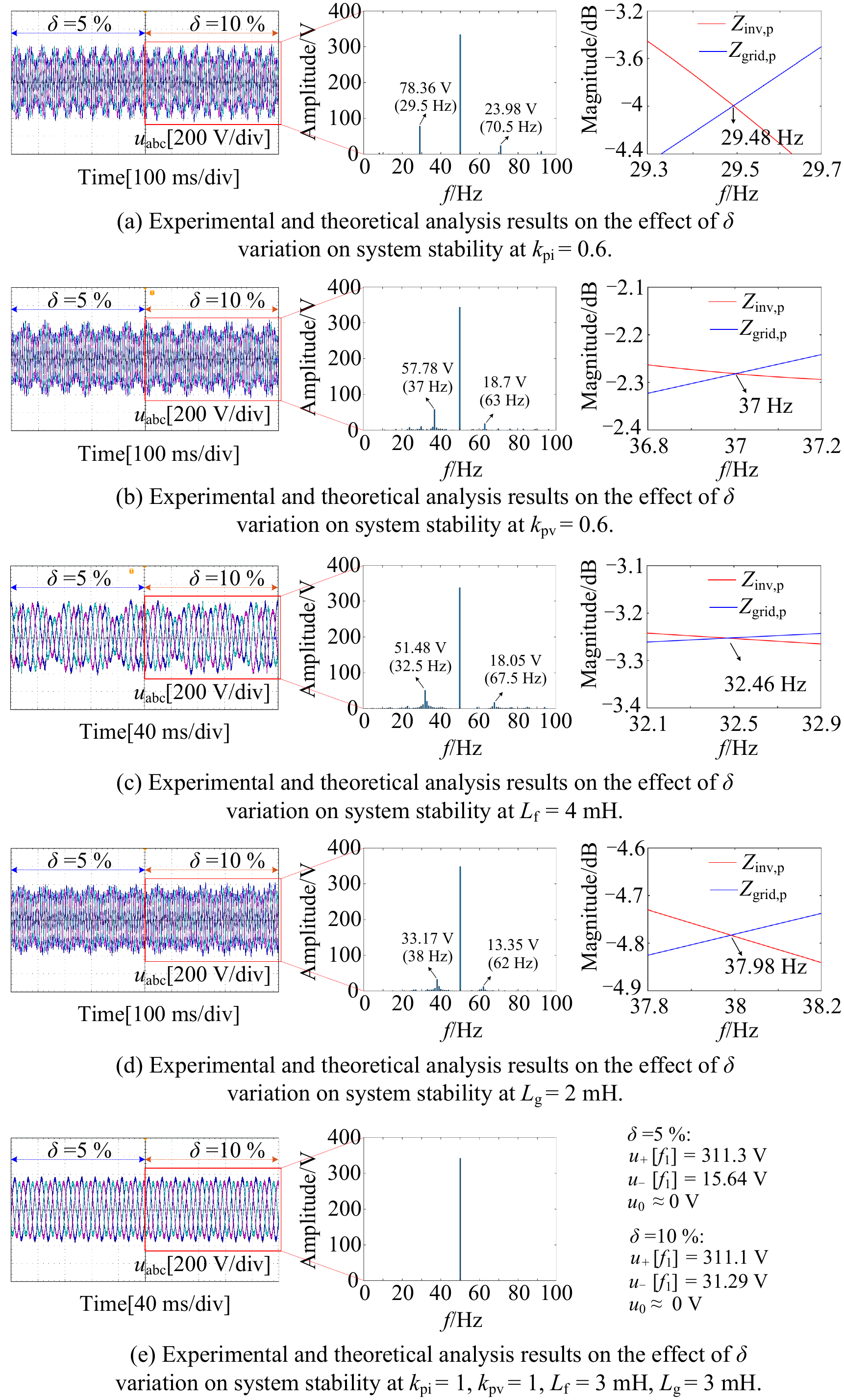}
	\caption{Experimental and theoretical analysis result of the effect of \(\delta\) variation on system stability.}
	\label{fig.11}
\end{figure}

Fig. 11 shows the experimental and theoretical analysis results of the effect of \(\delta\) variations on system stability. In Fig. 11(a)–(d), the three subplots from left to right represent the PCC voltage waveform, the FFT result of the captured waveform, and the theoretical analysis of the oscillation frequency. It can be observed that the system becomes unstable under the parameter conditions shown in Figs. 11(a)–(d), which agrees well with the results in Fig. 9. The oscillation frequency is defined as the frequency at which the magnitude curves of the DCI and grid-side equivalent positive-sequence impedances intersect, as shown in the third subplots of Fig.11(a)–(d). The coupling frequency is obtained by subtracting the oscillation frequency from twice the fundamental frequency. As shown in Fig. 11(a)–(d), FFT analysis of the captured experimental waveforms reveals both the oscillation frequency and the corresponding coupling frequency, which agree well with the theoretical predictions. This confirms that the proposed model accurately capture MFCE behavior of the DCI during oscillations. Furthermore, for the parameter setting in Fig. 11(e), the system remains stable, as verified by both the experimental waveforms and the FFT analysis results. The magnitudes of the fundamental positive-, negative-, and zero-sequence voltages for \(\delta\) = 5\% and \(\delta\) = 10\% are shown on the right-hand side.

\section{Conclusion}
This paper focuses on the droop-controlled grid-connected
systems under unbalanced conditions and develops SISO impedance models for the DCI and the grid side with comprehensive consideration of three-phase unbalanced operation and mirror frequency coupling effects. A comparison with existing DCI impedance models shows that the proposed model achieves higher accuracy. Additionally, the influence of the grid voltage unbalance on the MFCE and impedance characteristics of the DCI is investigated. Finally, the influence of dominant parameters on the system's stability margin is analyzed under three typical unbalanced conditions, providing theoretical guidance for parameter adjustments to improve system stability under unbalanced operating scenarios. The major conclusions are drawn as follows.

\begin{enumerate}
\item{The equivalent positive-sequence impedance of the DCI exhibits a capacitive negative resistance region in the sub/super-synchronous frequency ranges, making it susceptible to interaction-induced oscillations with the inductive grid impedance. In contrast, without such a capacitive negative resistance region, the negative-sequence system results in relatively better stability.}
\item{Grid voltage unbalance mainly affects the strength of the MFCE as well as DCI's equivalent positive-sequence impedance characteristics near the fundamental frequency.} As the unbalance level increases, its impact on the magnitude of the DCI's equivalent positive-sequence impedance becomes more significant, while its effect on the phase angle remains negligible.
\item{The effects of grid voltage and load unbalance on system stability are limited, whereas grid impedance unbalance has a much stronger impact. To ensure a sufficient stability margin, the feasible range of key parameters should be determined based on the phase margin threshold. In general, the proportional gains of the voltage and current control loops should be set to relatively high values, the filter inductance should be kept small, and the system should be designed to operate under weak-grid conditions.}
\end{enumerate}

The main attention of this work is paid to offline impedance modeling and stability analysis of DCIs under unbalanced operating conditions. Considering the time-varying conditions in practice, how to develop a more applicable online impedance modeling and stability analysis scheme will be investigated in the future. In particular, future research efforts may be devoted to incorporating data-driven techniques with theory-enabled analytical methods to fulfill this task. In addition, the modeling and stability analysis of droop-controlled inverters with positive- and negative-sequence decoupled control will be investigated in our future work.

\appendices

\bibliographystyle{IEEEtran}

\bibliography{Bibliography/mylib}

\begin{IEEEbiography}[{%
		\includegraphics[width=1in,height=1.25in,clip,keepaspectratio]{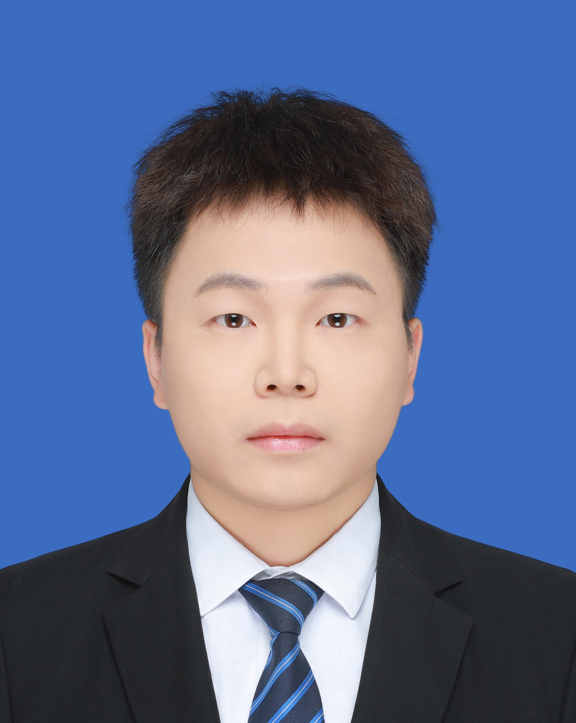}}]
	{Qiang Zeng} was born in Jiangxi, China, in 2000. He received the B.S. degree in electrical engineering from Nanchang University, China, in 2022. He is currently pursuing the Ph.D.\ degree in electrical engineering with the College of Electrical and Information Engineering, Hunan University, Changsha, China.
    
    His research interests include modeling and stability analysis of renewable-energy power systems.
\end{IEEEbiography}

\begin{IEEEbiography}[{
		\includegraphics[width=1in,height=1.25in,clip,keepaspectratio]{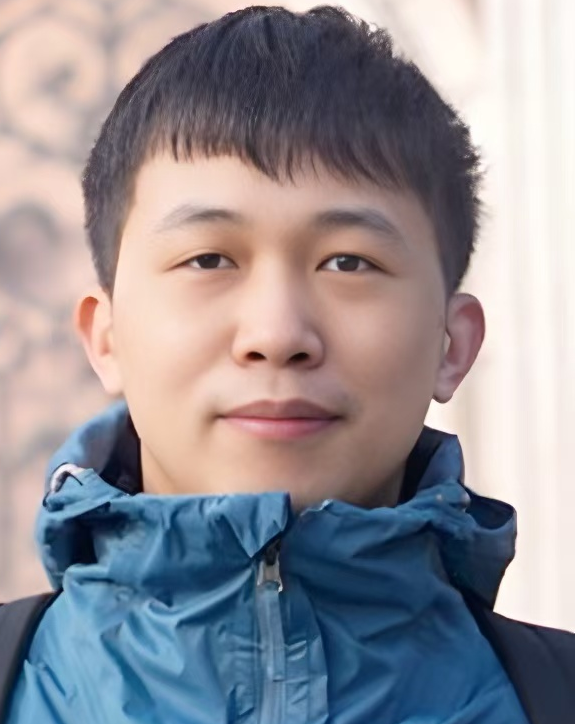}}]
	{Lipeng Zhu} (Senior Member, IEEE) received the B.S., M.S. and Ph.D. degrees from Huazhong University of Science and Technology, Wuhan, China, in 2012, from Wuhan University, Wuhan, China, in 2015, and from Tsinghua University, Beijing, China, in 2018, respectively, all in electrical engineering. He worked as a Post-Doctoral Fellow/Senior Research Assistant at The University of Hong Kong from 2018 to 2021. He is currently a Professor at Hunan University. His research interests mainly include data-driven power system stability and control, synchrophasor measurement technologies, spatial-temporal energy data analysis, and IoT data management and analytics in smart grids.
\end{IEEEbiography}

\begin{IEEEbiography}[{
		\includegraphics[width=1in,height=1.25in,clip,keepaspectratio]{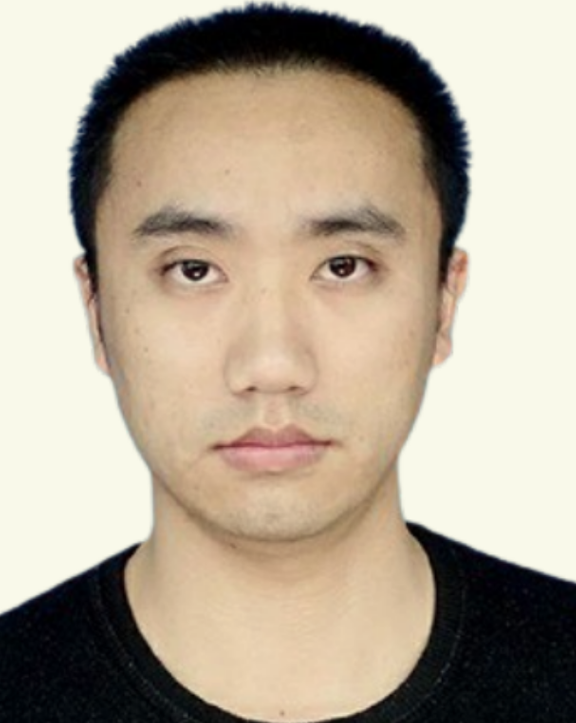}}]
	{Yang Li} received the B.S. degree in electrical engineering from the School of Electrical Engineering, Hebei University of Technology, Tianjin, China, in 2015, and the Ph.D. degree in electrical engineering from Hunan University, Changsha, China, in 2021.
	
	He is currently an Associate Professor with Hunan University. His research interests include power
	quality control and power electronics.
\end{IEEEbiography}

\begin{IEEEbiography}[{
		\includegraphics[width=1in,height=1.25in,clip,keepaspectratio]{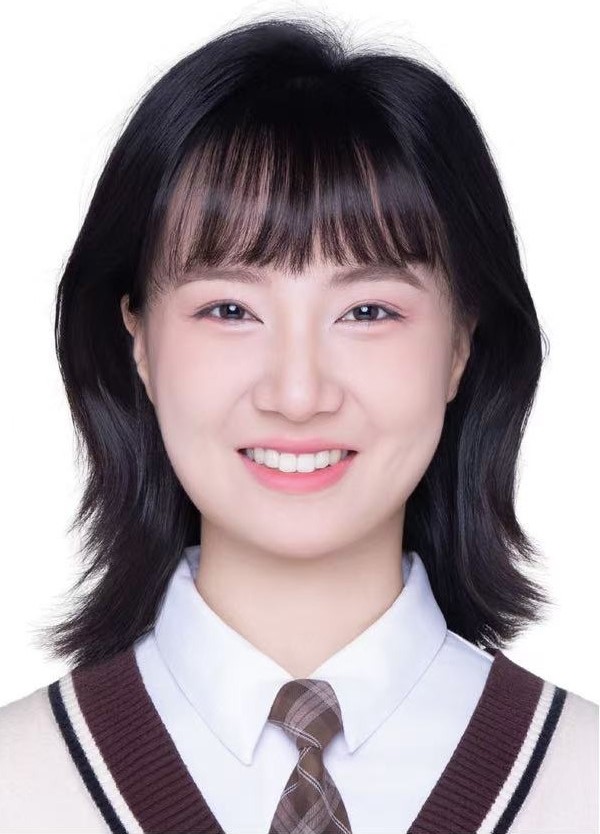}}]
	{Yi Lei} was born in Hunan, China, in 1999. She received the B.S. degree in electrical engineering
	from Beijing Jiaotong University, Beijing, China, in 2021. She is currently pursuing the Ph.D. degree in
	electrical engineering with the College of Electrical and Information Engineering, Hunan University,
	Changsha, China.
	
	Her research interests include power electronics, transient stability analysis and control, and cascaded
	DC power systems. 
\end{IEEEbiography}

\begin{IEEEbiography}[{
		\includegraphics[width=1in,height=1.25in,clip,keepaspectratio]{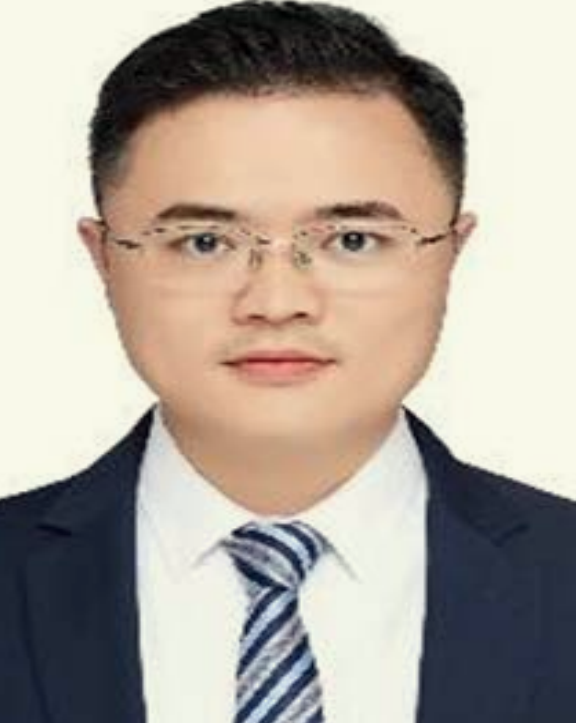}}]
	{Quan Zhou} (Senior Member, IEEE) received the B.S. and M.S. degrees from the department of electrical engineering, Shanghai Jiao Tong University, Shanghai, China, in 2011 and 2016, respectively. He received the Ph.D. degree from Illinois Institute of Technology, Chicago, IL, USA, in 2019. He is currently a full professor in the college of electrical and information engineering at Hunan University, Changsha, China.
\end{IEEEbiography}

\begin{IEEEbiography}[{
		\includegraphics[width=1in,height=1.25in,clip,keepaspectratio]{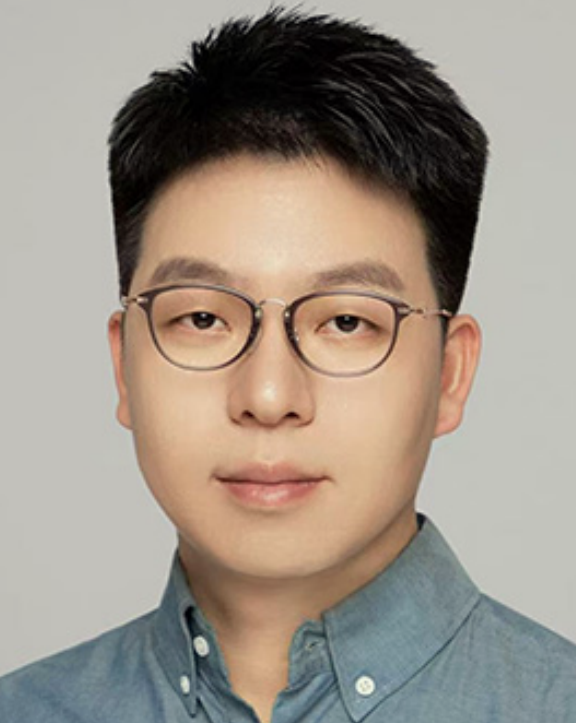}}]
	{Jiayong Li} (Senior Member, IEEE) received the B.Eng. degree from Zhejiang University, Hangzhou, China, in 2014, and the Ph.D. degree from The Hong Kong Polytechnic University, Hong Kong, in 2018. He is currently an associate Professor with the College of Electrical and Information Engineering of Hunan University. His research interests include distribution system planning and operation, power economics, demand-side energy management, and distributed control.
\end{IEEEbiography}

\begin{IEEEbiography}[{
		\includegraphics[width=1in,height=1.25in,clip,keepaspectratio]{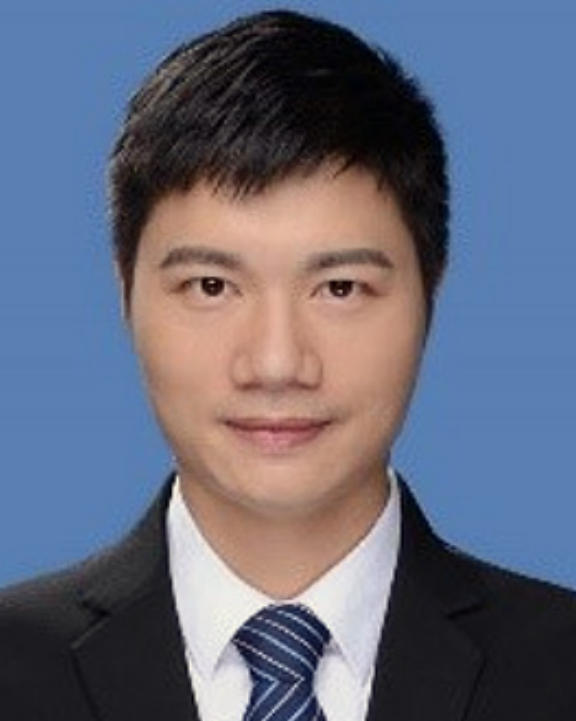}}]
	{Cong Zhang} (Senior Member, IEEE) received the D. degree in mathematics from the school of mathematics in South China, University of Tehnology, Guangzhou, China, in 2013. After that, he began to research electrical engineering, and he is currently an Associate Professor with the College of Electrical and Information Engineering, Hunan University, Changsha, China. His research interests include reactive power optimization, incorporating uncertainties, and interval power flow analysis.
\end{IEEEbiography}

\begin{IEEEbiography}[{
		\includegraphics[width=1in,height=1.25in,clip,keepaspectratio]{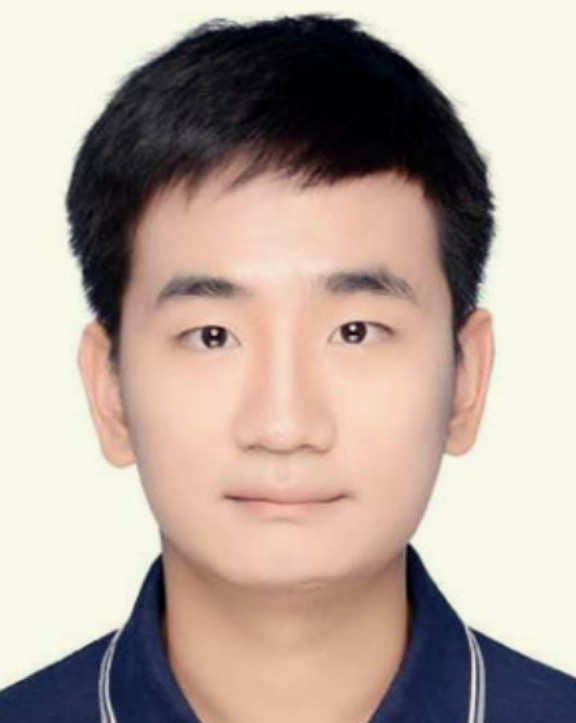}}]
	{Bingxu Li} was born in Inner Mongolia, China, in 1995. He received the B.Sc. degree in electrical engineering and automation from Xi’an Jiaotong University, Xi’an, China, in 2017. Since August 2019, he has been working toward the Ph.D. degree in electrical and electronic engineering with Nanyang Technological University, Singapore. 
	
	His research interests include optimization and control of heating, ventilation, and air conditioning (HVac) systems, fault detection and diagnosis technologies in HVac systems, smart buildings and building energy management systems.
\end{IEEEbiography}

\begin{IEEEbiography}[{
		\includegraphics[width=1in,height=1.25in,clip,keepaspectratio]{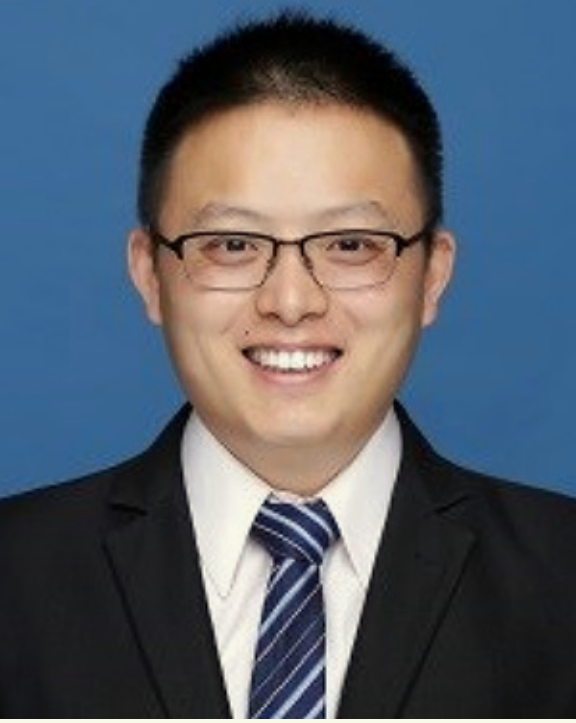}}]
	{Zhikang Shuai} (Senior Member, IEEE)  received the B.S. and Ph.D. degrees in electrical engineering from the College of Electrical and Information Engineering, Hunan University, Changsha, China, in 2005 and 2011, respectively. He was with Hunan University as an Assistant Professor from 2009 to 2012, an Associate Professor in 2013, and a Professor in 2014. His research interests include power quality control, power electronics, and microgrid stability analysis and control. Dr. Shuai was a recipient of the 2010 National Scientific and Technological Awards of China, the 2012 Hunan Technological Invention Awards of China, and the 2007 Scientific and Technological Awards from the National Mechanical Industry Association of China.
\end{IEEEbiography}

\end{document}